\newcommand{\cnotenum}{\texorpdfstring{CNOT}{CNOT}}

\documentclass[times,twocolumn,final]{elsarticle}

\usepackage{fix-cm}

\usepackage{medima}
\usepackage{framed,multirow}
\usepackage{stackengine}

\usepackage{epsfig}
\usepackage{amsmath}
\usepackage{lipsum}
\usepackage{enumitem}
\usepackage[toc, page]{appendix}

\usepackage{amssymb}
\usepackage{latexsym}
\usepackage{array, makecell}
\usepackage{bm}
\usepackage{xspace}
\usepackage{nicematrix}
\usepackage{subfigure}
\usepackage[T1]{fontenc}
\usepackage{graphicx}
\usepackage{multirow}
\usepackage{url}
\usepackage{xcolor}
\usepackage[colorlinks = true,
            linkcolor = tealblue,
            urlcolor  = tealblue,
            citecolor = tealblue,
            anchorcolor = tealblue]{hyperref}

\definecolor{newcolor1}{rgb}{.0, .502, .675}
\definecolor{candyAppleRed}{RGB}{255 8 0}

\makeatletter
\AtBeginDocument{\def\@citecolor{newcolor1}}
\AtBeginDocument{\def\@linkcolor{newcolor1}}
\AtBeginDocument{\def\@anchorcolor{newcolor1}}
\AtBeginDocument{\def\@filecolor{newcolor1}}
\AtBeginDocument{\def\@urlcolor{newcolor1}}
\AtBeginDocument{\def\@menucolor{newcolor1}}
\AtBeginDocument{\def\@pagecolor{newcolor1}}
\makeatother

\newcolumntype{?}{!{\vrule width 1pt}}
\usepackage{color, colortbl}
\definecolor{Gray}{gray}{0.9}
\definecolor{newcolor}{rgb}{.8,.349,.1}

\newif\ifcb
\cbtrue

\journal{}

\begin{document}

\verso{Chen \textit{et~al.}}

\begin{frontmatter}

\title{Deep learning-derived arterial input function for dynamic brain PET}

\author[1]{Junyu \snm{Chen}\corref{cor1}\fnref{fn1}}
\author[1,3]{Zirui \snm{Jiang}\fnref{fn1}}
\fnref{fn1}
\fntext[fn1]{Contributed equally to this work.}
\author[2,5]{Jennifer M. \snm{Coughlin}}
\author[4]{Ian \snm{Cheong}}
\author[4]{Kelly A. \snm{Mills}}
\author[1,6]{Martin G. \snm{Pomper}}
\author[1]{Yong \snm{Du}}
\address[1]{Department of Radiology and Radiological Science, Johns Hopkins Medical Institutions, Baltimore, MD, USA}
\address[2]{Department of Psychiatry and Behavioral Sciences, Johns Hopkins Medical Institutions, Baltimore, MD, USA}
\address[3]{Department of Biomedical Engineering, Johns Hopkins University, Baltimore, MD, USA}
\address[4]{Department of Neurology, Johns Hopkins University, Baltimore, MD, USA}
\address[5]{Department of Psychiatry, University of Texas Southwestern Medical Center, Dallas, TX, USA}
\address[6]{Department of Radiology, University of Texas Southwestern Medical Center, Dallas, TX, USA}
\cortext[cor1]{Corresponding author. E-mail address: 
 jchen245@jhmi.edu.}
\received{xxxx}
\finalform{xxxx}
\accepted{xxxx}
\availableonline{xxxx}
\communicated{xxxx}

\begin{abstract}
Dynamic positron emission tomography (PET) imaging combined with radiotracer kinetic modeling is a powerful technique for visualizing biological processes in the brain, offering valuable insights into brain functions and neurological disorders such as Alzheimer's and Parkinson's diseases. Accurate kinetic modeling relies heavily on the use of a metabolite-corrected arterial input function (AIF), which typically requires invasive and labor-intensive arterial blood sampling. While alternative non-invasive approaches have been proposed, they often compromise accuracy or still necessitate at least one invasive blood sampling. In this study, we present the deep learning-derived arterial input function (DLIF), a deep learning framework capable of estimating a metabolite-corrected AIF directly from dynamic PET image sequences without any blood sampling. We validated DLIF using existing dynamic PET patient data. We compared DLIF and resulting parametric maps against ground truth measurements. Our evaluation shows that DLIF achieves accurate and robust AIF estimation. By leveraging deep learning's ability to capture complex temporal dynamics and incorporating prior knowledge of typical AIF shapes through basis functions, DLIF provides a rapid, accurate, and entirely non-invasive alternative to traditional AIF measurement methods.
\end{abstract}

\begin{keyword}
\KWD Arterial Input Function \sep Dynamic PET
\end{keyword}

\end{frontmatter}



\section{Introduction}

Dynamic positron emission tomography (PET) is a molecular imaging technique involving the acquisition of sequential PET images over time, following radiotracer injection. By analyzing these dynamic PET data through kinetic modeling, one can quantify key physiological and biochemical parameters, including tissue receptor density, tracer influx or trapping rates, and competitive interactions between endogenous and exogenous ligands~\citep{dimitrakopoulou2021kinetic,rahmim2019dynamic}. Consequently, dynamic PET coupled with kinetic modeling has emerged as a crucial approach for investigating various diseases, particularly neurological disorders.

\begin{figure*}[t]
\begin{center}
\includegraphics[width=0.99\textwidth]{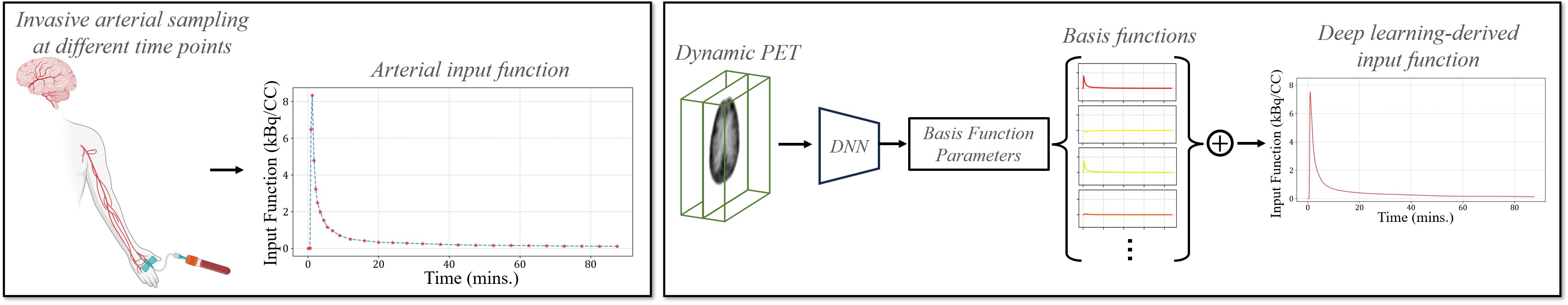}
\end{center}
   \caption{\texorpdfstring{The overview of the invasive arterial sampling used to obtain arterial input functions (left panel) versus the proposed non-invasive DLIF method (right panel).}{The overview of the invasive arterial sampling used to obtain arterial input functions (left panel) versus the proposed non-invasive DLIF method (right panel).}}
\label{fig:overview}
\end{figure*}

In kinetic modeling, sets of ordinary differential equations (ODEs) describe the dynamic relationships between a radiotracer, its physiological states, and the resulting PET images~\citep{gunn2001positron}. These tracer-specific models typically originate from compartmental frameworks in which the plasma concentration of the radiotracer over time serves as the input function~\citep{watabe2006pet}. The coefficients within these ODEs are solved using PET data, capturing the intrinsic kinetic properties of the tracer. Accurate determination of this input function is critical for robust kinetic modeling. Traditionally, this involves serial arterial blood sampling from the patient's radial artery, beginning at tracer injection (the start of dynamic PET scanning) and continuing until the scan concludes (as shown in the left panel of Fig. \ref{fig:overview}). Collected blood samples are centrifuged to isolate plasma, and the radiotracer concentration is measured using a well counter, with corrections applied for radiolabeled metabolites, thus producing a metabolite-corrected AIF~\citep{coughlin2018distribution}. Although serial arterial sampling is considered reasonably safe, dynamic PET scans typically last 90 to 180 minutes, and patient tolerance to arterial sampling can vary significantly, especially among elderly or medically compromised populations. This variability may lead to slower recruitment or increased dropout rates in longitudinal studies. Moreover, risks associated with arterial cannulation rise in aging patients, particularly those on medications that affect blood coagulation, potentially contraindicating arterial line placement~\citep{kang2018noninvasive}. Consequently, such concerns may deter researchers from adopting promising radiotracers in multicenter, longitudinal trials due to reluctance or the inability to perform arterial sampling. In addition, some PET research facilities may not have the capability to readily acquire dynamic, arterial measurements.

\textcolor{black}{Previous research has explored alternative, non-invasive methods for deriving the AIF, primarily focusing on population-based input functions (PBIF) and image-derived input functions (IDIF)~\citep{Zanotti_Fregonara_2012}. PBIF generates a standardized input function by averaging and normalizing arterial data across subjects, but it overlooks individual physiology, scanner differences, and acquisition quality, limiting accuracy~\citep{Zanotti_Fregonara_2012, Boutin_2007}. Additionally, PBIF cannot effectively distinguish between the parent radiotracer and its radiolabeled metabolites in the blood, typically requiring supplementary blood samples to perform accurate metabolite corrections~\citep{zanotti2011image, Takikawa_1993}. In contrast, IDIF estimates the whole-blood time-activity curve directly from dynamic PET images, capturing patient-specific variability for more individualized results. IDIF methods generally fall into segmentation- or statistical decomposition-based categories. Nonetheless, challenges such as accurate carotid segmentation, calibration via blood samples, and metabolite correction continue to hinder clinical adoption~\citep{zanotti2011image}.}

Recently, deep learning has emerged as a promising technique across various areas of medical imaging. Although still in its early stages, initial studies integrating deep learning methods have demonstrated improved accuracy AIF estimation, surpassing traditional IDIF and PBIF approaches~\citep{wang2020direct, ferrante2022physically,ferrante2024physically, Cui_2022, chen2023estimating, kuttner2024deep}. Our group previously pioneered the use of deep neural networks (DNNs) to directly estimate AIF from PET images~\citep{wang2020direct}. In this approach, a DNN processes two 3D PET volumes---one at a particular time point and another averaged over all time points---to predict the AIF value at each moment and subsequently reconstruct the full AIF curve. However, the method has limitations: since the network does not incorporate the entire dynamic PET sequence, it occasionally produces suboptimal estimates with non-smooth tails. \textcolor{black}{In contrast, \citet{kuttner2024deep} employed a DNN that uses the entire dynamic PET sequence as input to estimate AIF values at the corresponding time points. \citet{ferrante2022physically, ferrante2024physically} further advanced this direction by introducing a physically informed neural network (PINN)}, which estimates parameters of an analytical AIF model represented by a combination of two Gaussians and an exponential term modulated by a sigmoid function. Alternatively, \citet{Cui_2022} bypassed AIF estimation entirely, directly reconstructing parametric images using an unsupervised deep learning framework known as the conditional deep image prior. Nevertheless, these innovative methods remain constrained by limited sample sizes—stemming from inherent blood sampling difficulties—and by relatively low robustness when applied across different radiotracers.

In this study, we aim to overcome the limitations of current AIF estimation techniques and eliminate the need for manual intervention by developing a deep-learning-based method termed the \textit{deep-learning-derived input function} (DLIF). Building upon our preliminary work~\citep{wang2020direct,chen2023estimating}, DLIF directly estimates metabolite-corrected AIF from dynamic brain PET data. \textcolor{black}{Although conceptually related to the recent work by Ferrante~\textit{et~al.}~\citep{ferrante2022physically, ferrante2024physically}, which estimates analytical parameters of the Parker model, the proposed DLIF framework offers greater modeling flexibility by representing the input function through a composition of basis functions rather than enforcing a parametric functional form. This design enables DLIF to capture complex temporal dynamics beyond those expressible by a limited number of Gaussian or exponential terms.} Once trained, DLIF provides a fully non-invasive alternative, circumventing the challenges associated with traditional image segmentation methods and invasive arterial blood sampling. This advancement promises immediate benefits to ongoing and future research by improving patient comfort, significantly cutting operational costs related to anesthesiologist services, blood sampling, personnel, and facility logistics. Additionally, DLIF has the potential to enhance participant recruitment and retention in longitudinal studies.

\textcolor{black}{The remainder of the paper is organized as follows. Section~\ref{sec:rel_works} discusses related work. Section~\ref{sec:methods} describes the proposed methodology. The experimental setup, implementation details, and datasets used in this study are discussed in Sect.~\ref{sec:exp}. Section~\ref{sec:results} presents the experimental results. The findings drawn from these results are discussed in Sect.~\ref{sec:discussion}, and Sect.~\ref{sec:conclusion} concludes the paper. For clarity, all abbreviations used in this paper are summarized in Appendix~\ref{tab:list_abbr}.}

\section{Related work}
\label{sec:rel_works}
\subsection{Segmentation-based IDIF methods}
Segmentation-based IDIF relies on identifying and segmenting large blood vessels, such as segments of the aorta or femoral arteries, to estimate the whole-blood time-activity curve for individual patients. These vessels are chosen primarily due to their large size, which facilitates effective correction of partial volume effects (PVE)~\citep{Zanotti_Fregonara_2012}. However, in brain PET studies, the small size of cerebral vessels often results in significant PVE, substantially decreasing the accuracy and precision of the derived IDIFs~\citep{Zanotti_Fregonara_2012}. To address this challenge, numerous segmentation methods have been proposed, utilizing either PET images alone or PET images co-registered with MRI~\citep{litton1997input, chen1998noninvasive, wahl1999regions,liptrot2004cluster, parker2005graph, su2005quantification, mourik2008partial, fung2009multimodal, lee2012extraction, fung2013cerebral, sari2017estimation, khalighi2018image}. \textcolor{black}{Statistical decomposition approaches have further been introduced to mitigate PVE and noise. For example, \citet{chen1998noninvasive} modeled each voxel signal in the carotid region-of-interest (ROI) as a mixture of vascular and spillover activity:}
\begin{equation}
\label{eqn:chen_IDIF}
    c^{mea}(t) = rc\times c_p(t)+sp\times c_t(t),
\end{equation}
where $c^{mea}(t)$ denotes the measurement at time $t$ from the defined carotid artery ROI, $c_p(t)$ denotes the actual radioactivity from the artery, and $c_t(t)$ represents the radioactivity from the surrounding tissue at the same time point. The coefficients $rc$ and $sp$ refer to the recovery and spillover coefficients, respectively, which account for the respective contributions of the plasma and tissue radioactivities to the measured signal. In practice, $c^{mea}$ and $c_t$ are derived from dynamic PET data, using the carotid artery and adjacent tissue ROIs defined manually, whereas $c_p$ is estimated through venous blood samples collected at a few time points. Once $c^{mea}$, $c_p$, and $c_t$ are known, the least squares was employed to solve for $rc$ and $sp$.

\textcolor{black}{Subsequent work removed the need for invasive sampling. \citet{su2005quantification} applied ICA to the early dynamic PET frames and automated segmentation using Gaussian thresholding and dilation, while \citet{parker2005graph} incorporated PCA and graph-based Mumford-Shah segmentation to refine carotid delineation~. More recent efforts further improved artery segmentation through multimodal imaging and advanced algorithms~\citep{fung2009multimodal, khalighi2018image, su2013noninvasive, sari2017estimation, fung2013cerebral}.}

\subsection{Direct estimation of IDIF}
\textcolor{black}{
Alternatively, IDIF can be estimated directly from voxel values in dynamic PET, avoiding explicit segmentation of the carotid arteries~\citep{naganawa2005extraction, naganawa2005omission, naganawa2008robust, b2006extraction, wang2006model}. Such approaches commonly rely on statistical decomposition techniques, particularly independent component analysis (ICA), to separate voxel signals into plasma and tissue components. For example, Naganawa~\emph{et al.} proposed the following model to describe voxel values in dynamic PET images~\citep{naganawa2005extraction, naganawa2005omission}:
\begin{equation}
\label{eqn:naganawa_IDIF}
x(q,t)=s_p(q)c_p(t)+s_t(q)c_t(t),
\end{equation}
where $s_p(q)$ and $s_t(q)$ represent the plasma and tissue contributions of the voxel $q$, respectively, while $c_p(t)$ and $c_t(t)$ denote the uncalibrated IDIF and the tissue time activity curve (tTAC) at time $t$, respectively. When expressed in matrix form, Eqn. \ref{eqn:naganawa_IDIF} transforms into $\mathbf{X}=\mathbf{CS}$, where a modified ICA method known as EPICA decomposes the dynamic PET data matrix $\mathbf{X}$ into component matrices $\mathbf{C}$ and $\mathbf{S}$. The plasma-related component in $\mathbf{S}$ provides the uncalibrated IDIF. Similarly, Bodvarsson~\emph{et al.} employed non-negative matrix factorization (NMF) to derive the IDIF~\citep{b2006extraction}. Later, \citet{naganawa2008robust} advanced the intersectional searching algorithm (ISA)~\citep{wang2006model}, developing EPISA, which estimates a time-integrated IDIF directly from PET images for Logan graphical analysis~\citep{logan1990graphical}, using tTACs obtained from intensity-based clustering. A key limitation of decomposition-based methods is that they often cannot inherently determine the scale or sign of estimated components, requiring calibration with a single blood sample, typically taken at the AIF peak.}

\begin{figure*}[t]
\begin{center}
\includegraphics[width=0.95\textwidth]{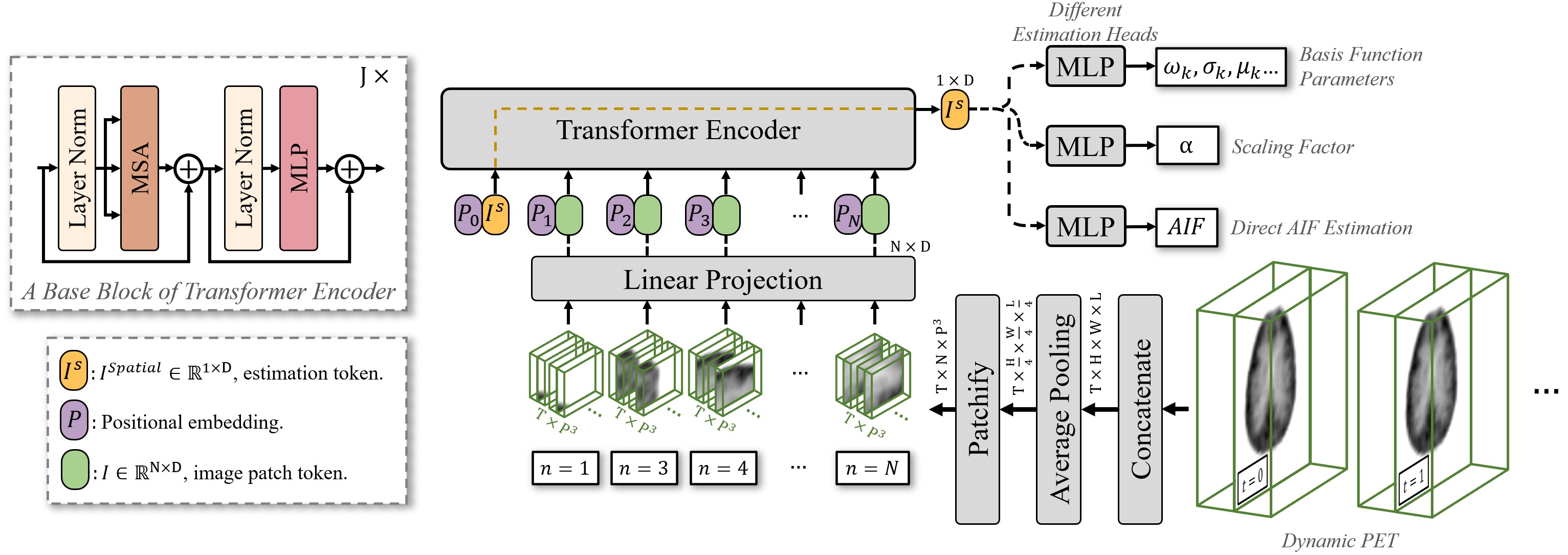}
\end{center}
   \caption{Overview of the proposed DLIF framework for AIF estimation. The model uses a ViT backbone with interchangeable AIF heads that either estimate the parameters of predefined basis functions to reconstruct the AIF or directly predict the full AIF curve.}
\label{fig:net_arch}
\end{figure*}

\section{Methods}
\label{sec:methods}
Let $I\in\mathbb{R}^{T\times H\times W\times L}$ be the skull-stripped dynamic PET image series, where $T$ represents the temporal dimension and $H\times W\times L$ defines the spatial dimension. Echoing previous studies on IDIF estimation, as discussed in the Introduction section, we model the voxel values in $I$ as a combination of contributions from both whole blood and tissue. This is expressed mathematically as:
\begin{equation}
\label{eqn:voxel_model}
\begin{split}
    I(t, \pmb{q}) &= f_b(\pmb{q})\Big[\alpha(\pmb{q}) c_p(t)+\beta(\pmb{q}) c_r(t)\Big]+f_t(\pmb{q})c_t(t)\\
\end{split}
\end{equation}
where $f_b(\pmb{q})$ and $f_t(\pmb{q})$ represent the contribution factors from whole blood and tissue, respectively, at voxel position $\pmb{q}$, and $\alpha(\pmb{q})$ and $\beta(\pmb{q})$ represent the respective factors from blood plasma and the rest of the blood components. Here, $\pmb{q}=(q_x, q_y, q_z)$ is the voxel index. The terms $c_p(t)$, $c_r(t)$, and $c_t(t)$ correspond to the plasma time activity curve or the metabolite-corrected AIF, the time activity curve in the rest of the blood (including radioactive metabolites and radioactivities in blood cells), and the tissue time activity curve (tTAC) at time $t$, respectively.

Our goal is to estimate the metabolite-corrected AIF or the DLIF, $c_p$, directly from the image sequence $I$ by employing deep learning. As illustrated in the Fig. \ref{fig:net_arch}, we first reduce the size of PET images by spatially averaging each dimension by a factor of 4, resulting in an image size of $T\times\frac{H}{4}\times\frac{W}{4}\times\frac{L}{4}$. The primary purpose of this downsampling is to reduce the computational burden. It is worth noting that estimating $c_p$, which, as shown in Eqn. \ref{eqn:voxel_model}, is not dependent on the voxel location $\pmb{q}$, and considering that neighboring voxels usually exhibit similar kinetic models, this size reduction theoretically should not affect the accuracy of the estimation.

To achieve our goal, we introduce a novel DNN framework. This network is built on top of ViT~\citep{dosovitskiy2021an}, enhanced with additional modules customized to process time-sequenced image volumes, such as dynamic PET scans. The output of ViT consists of a set of parameters that define the shapes of basis functions, which are then used to model the DLIFs.

\subsection{Network architecture}
\label{sec:net_arch}
\subsubsection{Vision Transformer}
\label{sec:dlif}
Transformers, initially developed for natural language processing tasks~\citep{vaswani2017attention}, have demonstrated significant potential in computer vision tasks~\citep{dosovitskiy2021an, liu2021swin, han2022survey}. Following these successes, they have been increasingly applied in the processing of medical images~\citep{li2023transforming}. A notable advantage of Transformers is their scalability~\citep{zhai2022scaling, liu2022convnet} and their capacity to capture long-range relationships between parts of the input. The decision to choose ViT over convolutional neural networks (ConvNets) for our application of AIF estimation is based on clear reasoning. ConvNets typically adopt small convolution kernels, usually ranging in size from 3 to 7, which inherently assume strong correlations between adjacent points (i.e., inductive bias). This presents two drawbacks for our application. First, AIF estimation often involves comparing spatially distant voxels, as seen in ISA~\citep{wang2006model} and EPISA~\citep{naganawa2008robust} described in the Introduction section. Localized convolution kernels in ConvNets are not as effective in capturing these distant relationships, which only become implicit as the network layers deepen. Second, for precise estimation across the temporal dimension, it is preferable to process the entire temporal data directly, rather than applying small convolution kernels across time points. It is important to acknowledge recent advancements that show that ConvNets can be optimized with much larger kernels~\hbox{\citep{ding2022scaling, liu2023more}} or advanced convolutional operations~\citep{liu2022convnet}, potentially matching the capabilities of Transformers. However, optimizing the network architecture to balance Transformers and ConvNets is beyond the scope of our study. In this work, we have benchmarked the proposed network against commonly used DNNs, with further details and results presented in the Results section. 

The proposed network is built on the foundation of ViT~\citep{dosovitskiy2021an}, which was originally designed for 2D image processing. Recognizing the limitations of ViT in this regard, we have expanded its functionality to effectively handle 3D image volumes. \textcolor{black}{The architecture of the neural network is depicted in Fig.~\ref{fig:net_arch}, and the details regarding the ViT and the associated attention mechanism are described in \ref{sec:vit}}

\subsubsection{Estimation head}
In the original ViT~\citep{dosovitskiy2021an}, the classification head is attached to the class token for the final prediction. In a similar fashion, our model attaches different estimation heads to the estimation token, $I^\text{spatial}\in\mathbb{R}^{1\times D}$, \textcolor{black}{as shown in Fig.~\ref{fig:net_arch}}. Since this token engages with all others during self-attention, it is capable of learning the global information necessary for AIF estimation. The estimation heads are responsible for producing either a direct AIF estimation, which has an output dimension of $1\times T$, or a set of parameters to combine the basis functions. The latter is discussed in the following section.

\subsection{Basis functions}
\label{sec:basis_func}
Given that AIFs generally exhibit certain smoothness characteristics, such as typically smoother tails, employing basis functions for their parameterization is beneficial for maintaining this smoothness in the estimations. An additional benefit of employing basis functions, as opposed to direct AIF estimation, is their ability to provide continuous functions. Direct estimation, in contrast, can only yield values at discrete time points and may require potentially error-prone interpolation to fill in values between these points. In our study, we explored the use of Gaussian and truncated exponential functions as basis functions. These have historically been used for modeling AIFs due to their inherent properties that align well with the characteristics of AIFs. Gaussian functions, for instance, have been employed in \citep{mlynash2005automated, parker2006experimentally}, while exponential functions have seen use in \citep{feng1993models, parsey2000validation}. However, unlike these traditional methods where the number, scale, and sign of the basis functions are predetermined, our method capitalizes on the complex modeling capabilities of DNNs. This allows the DNN to freely learn and adjust all parameters of the basis functions to achieve the most accurate fit.

The proposed DLIF framework produces $K$ sets of parameters corresponding to $K$ basis functions. In theory, if $K$ equals $T$ (i.e., the number of time points), the AIF values at each of these $T$ points can be accurately represented using $K$ Kronecker delta functions. This can be achieved by aligning the amplitude of these functions with the AIF values and minimizing the width of the basis functions. Consequently, having $K\leq T$ is considered adequate for providing a precise estimation of the AIF. In the following paragraphs, we detail the basis functions that have been considered in this study.

\subsubsection{Gaussian basis function}
The superposition of $K$ Gaussian functions can be expressed mathematically as:
\begin{equation}
    \hat{DLIF} = \sum_{k=0}^{K-1}\frac{\omega_k}{\sigma_k\sqrt{2\pi}}\exp\bigg({-\frac{(t-\mu_k)^2}{2\sigma_k^2}}\bigg),
\end{equation}
where $\sigma_k$ and $\mu_k$ determine the scale and location of the $k$-th Gaussian function, respectively, and $\omega_k$ specifies the weight of the $k$-th Gaussian function.

\subsubsection{Exponential-sigmoid Basis Function}
When using exponential functions for modeling AIF, it is important to consider that these functions do not inherently reduce to zero. Yet, the AIF value prior to the injection of the radioactive tracer is in fact zero. To accommodate this, previous methods have employed a truncated exponential model that introduces a discontinuity for the AIF~\citep{feng1993models, parsey2000validation}. In such models, the estimated IDIF is set to zero before a specified time point. After this time point, the IDIF is then modeled to follow a form of exponential functions, aligning with the behavior observed pre- and post-tracer injection. However, the discontinuity inherent in previous AIF models using exponential functions poses a challenge, as it is not differentiable and thus prevents the backpropagation of gradients needed to update DNN parameters during training. To address this, we have devised a workaround by relaxing the discontinuity with a sigmoid function, as similarly done in by \citep{parker2006experimentally}. In our model, the superposition of $K$ exponential-sigmoid functions is formulated as:
\begin{equation}
    \hat{DLIF} = \sum_{k=0}^{K-1}\omega_k\lambda_k\exp\bigg(-\lambda_k(t-\gamma_k)^2\bigg)\cdot\frac{1}{1+\exp(-\eta_k(t-\gamma_k))},
\end{equation}
where the first component represents the exponential function, and the second part embodies the sigmoid function. The center of each exponential-sigmoid function, represented by the $k$-th function, is located at $\gamma_k$. The scale and weight of these functions are characterized by $\lambda_k$ and $\omega_k$, respectively. Additionally, $\eta_k$ is the parameter that controls the steepness of the sigmoid function.

The parameters for these basis functions are produced by the proposed DLIF framework. For both types of basis functions, the scale parameters undergo a ReLU activation function, followed by an addition of $\epsilon$, a small value introduced to prevent the scale parameters from assuming values equal to zero. In contrast, the other parameters are not subjected to any activation functions, allowing them the flexibility to assume any value as needed.

\subsection{Scaling factor estimation head}
We take an additional step to differentiate the estimation of the AIF's shape from its amplitude. This is achieved by incorporating a separate head in our DNN, dedicated to estimating a scaling factor $\alpha$. This factor adjusts the estimated IDIFs ($\hat{DLIF}$) to align with the amplitudes of the true AIFs:
\begin{equation}
    DLIF = \alpha\cdot\hat{DLIF}.
\end{equation}
This echoes the approach commonly used in traditional decomposition-based IDIF estimation techniques~\citep{naganawa2005extraction, naganawa2005omission, b2006extraction}, where blood sampling at the peak of AIF is used to scale the IDIF. However, in our model, the scaling factor is determined by the DNN. Through empirical evaluation, we have found that this approach of using a scaling factor yields better results than having the DNN directly output the IDIF without any scaling. The evidence supporting this finding is detailed in a later section of the paper.

\subsection{Loss function}
\subsubsection{AIF similarity loss function} To train the DNN, we employed an $\ell_1$-based loss function. The preference for $\ell_1$ loss over $\ell_2$ loss is driven by the need to assign equal penalty strength to the IDIF values at each time point. Since the mean AIF values are typically dominated by the peak, which is several orders of magnitude higher, using $\ell_2$ loss would disproportionately assign a larger penalty to the peak. This is undesirable, as it can lead to an unbalanced focus on matching the peak value at the expense of other time points. Therefore, $\ell_1$ loss is chosen to ensure a more balanced training. The loss is defined as:
\begin{equation}
    \mathcal{L}(DLIF, AIF)=\frac{1}{T}\sum_t^T|DLIF(t)-AIF(t)|.
\end{equation}

In the case of direct estimation, the predicted IDIF is directly compared with the true AIF. However, when employing basis functions, the values of the estimated IDIF are frist derived from the aggregated basis functions at the same time points as the true AIF. Then, a comparison is made between these sampled values and the true AIF.

\subsubsection{Sparsity constraint} The selection of the optimal number of basis functions is a critical hyperparameter that necessitates careful adjustment. To mitigate the need for this tuning and to simplify the AIF representation via basis functions, thus preventing overfitting, we implement a sparsity constraint on the basis functions' weights. This approach promotes sparse weight distributions, encouraging the majority of the weights to approximate zero. The constraint is formally defined as follows:
\begin{equation}
    \mathcal{R}(\omega)=\frac{1}{K}\sum_k^K|\omega_k|,
\end{equation}
where $K$ is the total number of basis functions employed, and $\omega_k$ represents the weight of the $k$-th basis function. This sparsity constraint enables the use of a larger number of basis functions while mitigating the risk of overfitting in the AIF modeling.

\section{Experiments}
\label{sec:exp}
\subsection{Dataset and preprocessing}
The dataset used in this work were de-identified dynamic [\textsuperscript{11}C]DPA-713 brain PET images. [\textsuperscript{11}C]DPA-713 is a second generation PET tracer targeting the translocator protein 18kDa (TSPO) for detecting microglial response or proliferation in vivo ~\citep{Chauveau_2008, Venneti_2006, Tichauer_2015, Muzi_2012, coughlin2014regional, Wang_2016}.  It has shown superior binding affinity and signal-to-background ratio \citep{Venneti_2006, Tichauer_2015, Muzi_2012, coughlin2014regional, Wang_2016}. The [\textsuperscript{11}C]DPA-713 affiliation is affected by single nucleotide polymorphism (rs6971) TSPO genotyping, with C/C, C/T 
 and T/T corresponding to high-affinity binders (HAB), mixed-affinity binders (MAB), and low-affinity binders (LAB), respectively.\citep{Owen201118kDa, Milenkovic2018Effects} The data used in this study were collected through several clinical research studies at Johns Hopkins University that used [\textsuperscript{11}C]DPA-713-TSPO-PET in health control individuals \citep{endres2009initial}, and several conditions that may cause neuroinflammation, such as HIV, Lyme disease, and repeated traumatic brain injury \citep{Coughlin_2015, Coughlin_2018, rubin2018microglial, rubin2023imaging, rubin2022imaging}. 
 
From these studies, de-identified data from HAB and MAB participants were collected, including 37 controls and 86 patients. Among those 67 were HAB and 56 were MAB. The data includes 90 minutes dynamic PET images and metabolite corrected AIF measured from blood sampling. The PET data were acquired on a brain-dedicated High Resolution Research Tomograph (HRRT, Siemens Healthcare, Knoxville, TN), with a fitted thermoplastic facemask for head fixation to reduce motion. The 90 min data were binned into 30 frames and reconstructed using the iterative ordered subsets expectation maximization algorithm \citep{Coughlin_2018}. The data were acquired from each participant through studies approved by the Johns Hopkins Institutional Review Board. Each participant provided written informed consent, which included the use of their data in secondary analyses. 
 
All dynamic PET images of the brain were first aligned with a template through affine registration. Subsequently, skull stripping was applied to exclude non-brain regions. This procedure was carried out using SynthStrip, a publicly available learning-based method~\citep{hoopes2022synthstrip}. In particular, SynthStrip was applied to an image averaged over all time frames, and the generated mask was then used to consistently strip the brain regions across the time frames. Subsequently, the images were uniformly cropped to a size of $160\times192\times160$, maintaining an isotropic resolution of 1 $mm$. Additionally, the uptakes in the images were normalized to the Standardized Uptake Value (SUV), and the input function was correspondingly normalized to ensure consistency in the data. For training and evaluation, we implemented five-fold cross-validation, with each fold comprising an 8:2 split for training and testing.

\subsection{Implementation details}
\label{sec:imp_details}
The models were implemented using the PyTorch framework~\citep{paszke2019pytorch} on a PC equipped with two NVIDIA Quadro P6000 GPUs and an NVIDIA RTX A4000 GPU. The training was conducted over 500 epochs, using the Adam optimizer~\citep{kingma2014adam} with a learning rate of 1e-4 and a batch size of 1. To enhance the robustness of the models, the dataset underwent augmentation through random spatial direction flipping during the training phase.

\subsection{Baseline methods}
To evaluate the effectiveness of our proposed method, we conducted comparisons with two popular convolutional neural networks, ResNet50~\citep{he2016deep} and ConvNeXt~\citep{liu2022convnet}, as well as a traditional IDIF estimation method, EPICA~\citep{naganawa2005extraction, naganawa2005omission}, that predicts AIF directly from dynamic PET image sequences. The specifics of these baseline methods are outlined below.

\begin{itemize}
\item \textit{ResNet50}---We first compare the proposed network architecture with a widely used ConvNet architecture, ResNet50~\citep{he2016deep}, which has found extensive applications across a diverse range of image classification and regression tasks. To ensure a fair comparison, the input configuration for ResNet50 mirrors that of our proposed network. This involves reducing the original spatial resolution of the dynamic PET sequence by a factor of 4 and concatenating it along the temporal axis, resulting in the formation of $T$ channels.

\item \textit{ConvNeXt}--- We then evaluate the proposed DLIF against \hbox{ConvNeXt}~\citep{liu2022convnet}, a recent ConvNet architecture that has been recognized as a strong counterpart to Transformer models. We employ the default configuration of ConvNeXt but convert its 2D modules into 3D modules with the input remaining consistent with the aforementioned specifications.

\item \textit{EPICA}---The final baseline method compared in this study is EPICA~\citep{naganawa2005extraction, naganawa2005omission}, a statistical decomposition-based method discussed in the Introduction section. EPICA requires a scaling factor, typically derived from a single blood sample, to calibrate the estimated IDIF to align with the amplitude of the actual AIF. In our evaluation, we used the peak value of the ground-truth AIF as the scaling factor for the estimated IDIF.
\end{itemize}

\subsection{Evaluation metrics}
The DLIFs generated by the proposed framework have been applied in Logan graphical analysis. However, it is important to note that the potential applications of the proposed DLIF are not confined to this specific analysis technique. Logan graphical analysis employs linear regression to analyze the data after a specified time, estimating the slope of the resulting line to determine the total volume of distribution, as discussed in~\citep{carson2005tracer}. In this context, an exact match between the DLIF and the actual AIF before this chosen time is not critical. The crucial aspect is that the integral of the DLIF (i.e., the area under its curve) matches that of the AIF prior to this time. After this time point, the curves should align as closely as possible. This precise alignment is vital for accurately estimating the slope, which is crucial for the correct calculation of the total volume of distribution. Given these considerations, using metrics such as the mean squared error or the mean absolute error to assess the accuracy of the DLIFs may not provide the most meaningful insights. This is because such metrics could be disproportionately influenced by the errors in matching the peak values, which are significantly higher in amplitude compared to the input function values at later time points.

\subsubsection{Pearson's correlation coefficient (\texorpdfstring{$\pmb{r}$}))} To evaluate the accuracy of AIFs estimated by various methods, we adopt Pearson's correlation coefficient or $\pmb{r}$, a metric also employed in related research on AIF estimation in perfusion CT~\citep{de2021aifnet}. This metric quantifies the correlation between the estimated AIF and the ground truth, and is mathematically defined as:
\begin{equation}
    \pmb{r}(\hat{y}, y)=\frac{\sum_t^T(\hat{y}_t-\Bar{\hat{y}})(y_t-\Bar{y})}{\sqrt{\sum_t^T(\hat{y}_t-\Bar{\hat{y}})^2}\sqrt{\sum_t^T(y_t-\Bar{y})^2}},
\end{equation}
where $\hat{y}, y$ represent, respectively, the estimated AIF and the ground truth AIF. Note that $\pmb{r}$ specifically quantifies the linear relationship between two variables, essentially measuring the similarity in the shape of two curves. However, $\pmb{r}$ does not assess the absolute matching of the amplitudes or the integrals of these curves.

\begin{figure}[t]
\begin{center}
\includegraphics[width=0.3\textwidth]{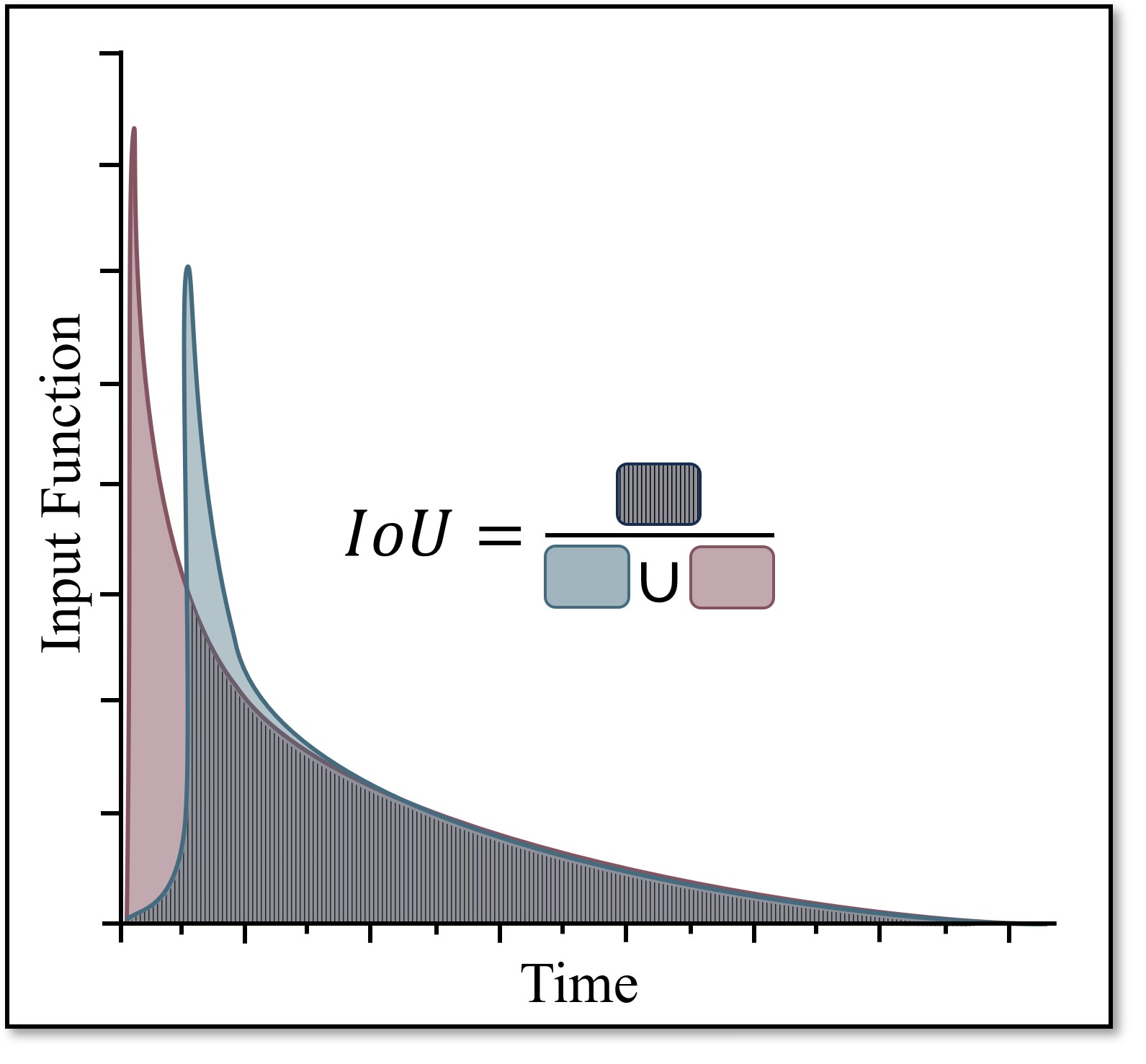}
\end{center}
   \caption{The computation of IoU between the estimated and the true AIFs.}
\label{fig:IoU}
\end{figure}

\begin{figure*}[t]
\begin{center}
\includegraphics[width=0.99\textwidth]{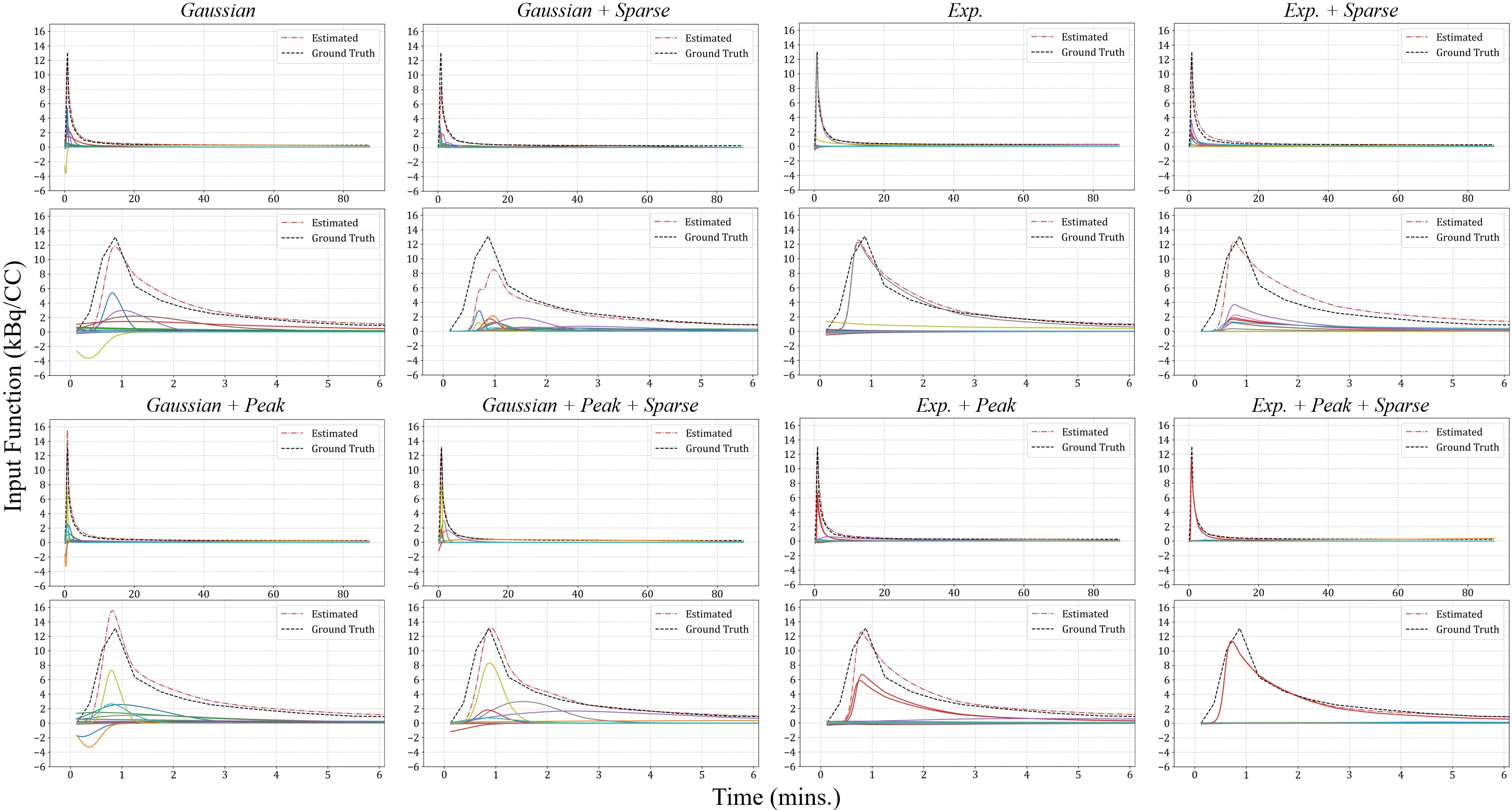}
\end{center}
   \caption{\textcolor{black}{This figure illustrates how various basis functions come together to assemble DLIFs for the same subject across different configurations. The first and third rows show the overall composition of the DLIFs for the eight models throughout the 90 minutes duration, whereas the second and fourth rows provide a closer look at the first 0 to 6-minute interval. Here, the red and the black dashed lines correspond to the estimated DLIFs and the actual ground truth AIFs, respectively. Meanwhile, the solid lines denote the individual basis functions.}}
\label{fig:DLIF_compose}
\end{figure*}

\subsubsection{Intersection over Union}
To complement $\pmb{r}$, we propose employing an additional metric, Intersection over Union (IoU), that evaluates the integrals or the areas under the curves of the two entities. Illustrated in Fig. \ref{fig:IoU}, the IoU metric we propose for evaluating AIFs is mathematically defined as:
\begin{equation}
    IoU(\hat{y}, y) = \frac{\sum_t^T \min(\hat{y}_t, y_t)}{\sum_t^T \max(\hat{y}_t, y_t)}.
\end{equation}
Although IoU is a metric frequently used in computer vision for measuring the overlap between two areas or volumes in image segmentation or detection tasks, we propose to adapt it here to measure the ratio of the overlap between two areas under the input function curves. Considering the pronounced sharpness typically seen in the peaks of AIFs, the IoU metric inherently places less emphasis on matching these peak points. Instead, IoU focuses more on the alignment of the overall amplitude and the cumulative area under the curve, which is more aligned with the accuracy needs of downstream Logan graphical analysis.

\subsubsection{Root mean squared error}
We also incorporated the widely recognized root mean square error (RMSE) to evaluate estimated AIF. Its mathematical formulation is:
\begin{equation}
    RMSE(\hat{y}, y) = \sqrt{\frac{1}{T}\sum_t^T(\hat{y}_t-y_t)^2}.
\end{equation}

\subsubsection{Logan graphical analysis}
The Logan graphical analysis is widely used for quantifying dynamic PET images of a reversible tracer, where the tissue time-activity curve (TAC) is mathematically transformed and plotted against ``normalized time'' to estimate the total distribution volume ($V_T$).  It is mathematically defined as:
\begin{equation}
     \frac{\int_{0}^{t} C_T(\tau) \, d\tau}{C_T(t)} = V_T \frac{\int_{0}^{t} C_P(\tau) \, d\tau}{C_T(t)} + \text{intercept}
\end{equation},
where $C_{T}(t)$ is the measured tissue activity and $C_p(t)$ is the input curve. For the reversible compartment, this expression results in a straight line with a slope of $V_T$ after an equilibration time (t*). In previous studies, we have demonstrated that Logan graphical analysis with a t* = 30 minutes could provide robust $V_T$ values that agree with those produced from compartmental models \citep{endres2009initial}.\newline \newline

Recognizing the possibility of bias in evaluation metrics, which may lean towards error in peak matching, as the peak value in an AIF usually overshadows other values, we considered additional steps in our analysis. Specifically, in Logan graphical analysis, the alignment of AIFs post-30 minutes is of greater significance than peak matching. To address this, we divided the AIFs into two segments: one before 30 minutes and one after. This division allowed us to separately assess and report the \textit{IoU} and \textit{RMSE} metrics for each segment, ensuring a more balanced and relevant assessment. Meanwhile, for $\pmb{r}$, which primarily measures the correlation or shape similarity between the estimated and true AIFs, we report the evaluation for the entire duration.

\begin{figure*}[t]
\begin{center}
\includegraphics[width=0.99\textwidth]{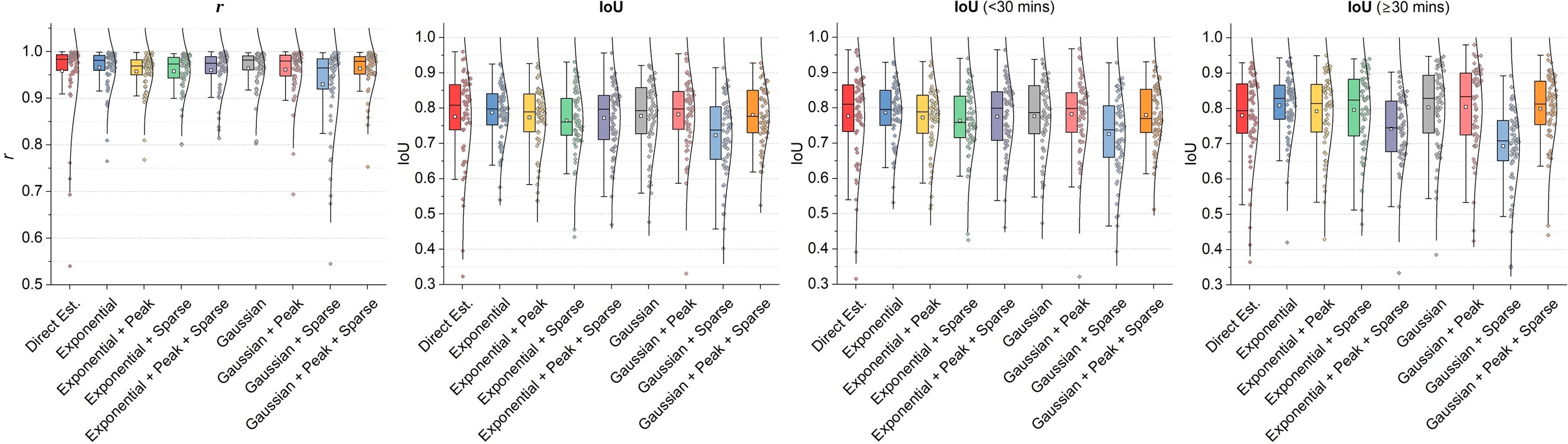}
\end{center}
   \caption{Comparative results on the validation dataset from the ablation study and component analysis. The first two graphs display the overall PCC and IoU values for the models under comparison. Subsequent graphs illustrate the IoU values compared within two time frames: before and after 30 minutes.}
\label{fig:ablation}
\end{figure*}

\subsubsection{Statistical Tests}
To statistically analyze and compare the performance of the proposed DLIF framework with baseline models, including EPICA, ResNet50, and ConvNeXt, we used the Wilcoxon signed rank test. This non-parametric test is commonly employed to compare the performance of machine learning models on paired samples, particularly when the data does not meet the assumption of normal distribution required for a t-test. We applied this statistical test to compare the top-performing model with the second-highest performing model among the baseline methods, including EPICA, ResNet50, and ConvNeXt. We conducted repeated tests for identical scores with Bonferroni correction to adjust the $p$-values for multiple comparisons.

Furthermore, we used the Mann-Whitney U test to evaluate the performance of the models in two different genotypes. This test is also non-parametric and is used on independent samples where the data does not meet the assumption of normal distribution, making it suitable for testing statistical significance between subjects of different genotypes.

\subsection{Ablation study and component analysis}
In pursuit of optimizing the proposed DLIF framework, we undertook an ablation study and component analysis. This process aimed to pinpoint the most effective AIF estimation strategy within our framework. Each evaluated model is described by the following:
\begin{itemize}
    \item ``\textit{Direct Est.}''---This model directly predicts AIF values at specific discrete intervals, in accordance with the AIF values acquired using arterial sampling.
    \item ``\textit{Exp.}''---This model generates parameters for the exponential-sigmoid basis functions, which are then superimposed to create a continuous estimate of the AIF.
    \item ``\textit{Exp. + Peak}''---In addition to generating parameters for the exponential-sigmoid basis functions, this model adds a scaling factor to adjust the amplitude of the estimated AIF. 
    \item ``\textit{Exp. + Sparse}''---In addition to generating parameters for the exponential-sigmoid basis functions, this model applies a sparsity constraint to encourage employing a smaller set of basis functions.
    \item ``\textit{Exp. + Peak + Sparse}''---This model not only generates parameters for the exponential-sigmoid basis functions and a scaling factor but also includes a sparsity constraint.
    \item ``\textit{Gaussian}''---This model outputs parameters for Gaussian basis functions, which are used to model the AIF as a sum of Gaussian curves.
    \item ``\textit{Gaussian + Peak}''---In addition to the Gaussian basis function parameters, this model includes a scaling factor to scale the amplitude of the estimated AIF. 
    \item ``\textit{Gaussian + Sparse}''---In addition to the Gaussian basis function parameters, this model adds a sparsity constraint on the Gaussian basis functions, aiming to reduce the number of basis functions required for representing the AIF.
    \item ``\textit{Gaussian + Peak + Sparse}''---This model generates parameters for the Gaussian basis functions, includes a peak scaling factor and applies a sparsity constraint during training.
\end{itemize} 

\begin{figure*}[thp]
\begin{center}
\includegraphics[width=0.99\textwidth]{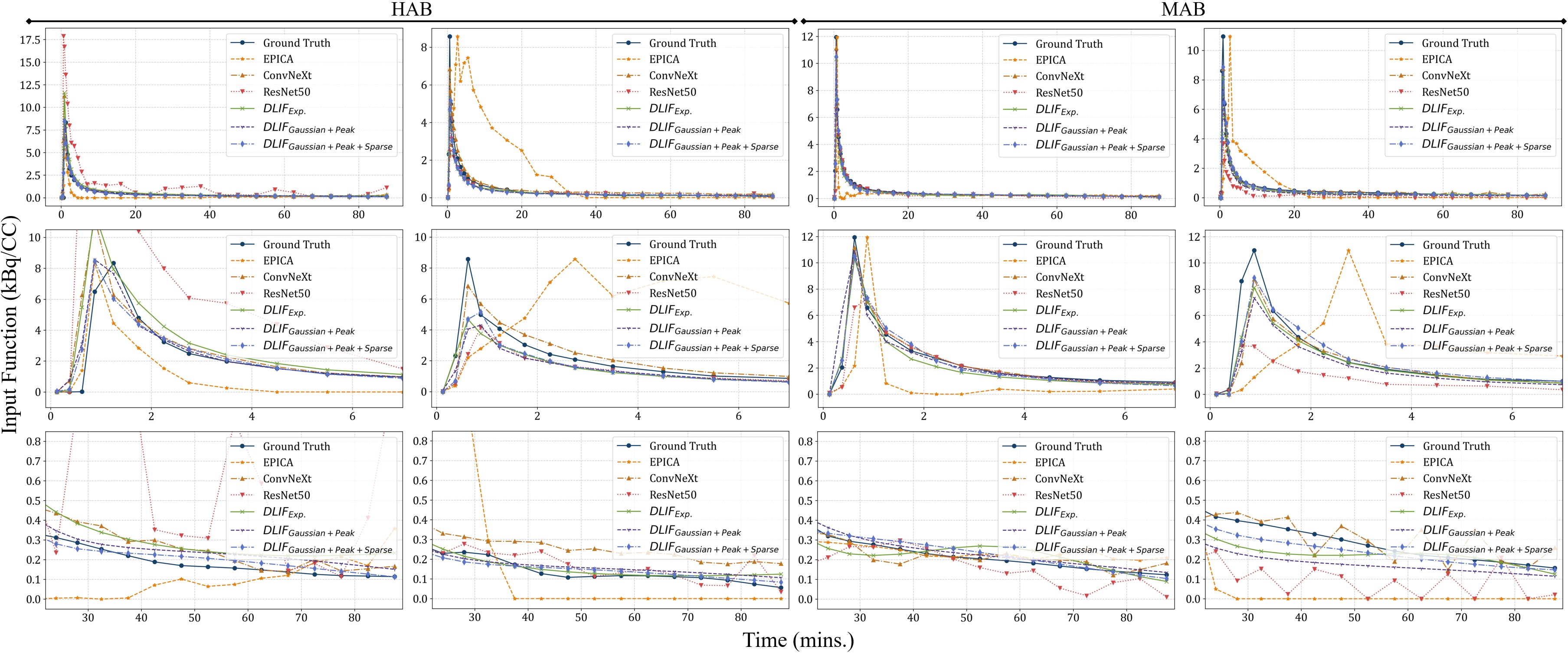}
\end{center}
   \caption{Qualitative comparison of baseline versus proposed techniques using five-fold cross-validation on both genotypes\textcolor{black}{, with each column representing results from a single patient.} The first two columns present results from two patients with HAB, contrasted with the last two columns showcasing results from MAB patients. The first row depicts overall estimated input functions, and the second and the third rows offer zoom-in views within the 0 to 7 minute and 22 to 90 minute timeframes, respectively.}
\label{fig:quali_res}
\end{figure*}

\section{Results}
\label{sec:results}
\subsection{Composition of DLIFs using basis functions}
We begin with a qualitative analysis of how basis functions combine within DLIFs across different configurations. Figure \ref{fig:DLIF_compose} illustrates the superposition process and highlights several key findings. 

First, configurations employing Gaussian basis functions typically activate a larger number of functions compared to those using exponential-sigmoid basis functions. Even when applying a sparsity constraint, multiple Gaussian functions remain active (e.g., ``\textit{Gaussian + Peak + Sparse}''), enabling DLIFs to accurately capture subtle details of the ground-truth AIF. Conversely, exponential-sigmoid configurations often utilize fewer basis functions. Notably, under sparsity constraints, the DNN can converge to solutions involving a minimal number of exponential-sigmoid functions---such as the ``\textit{Exp. + Peak + Sparse}'' configuration---where only one function remains active, effectively suppressing the others. This outcome highlights the intrinsic compatibility between the exponential-sigmoid shape and true AIF curves, facilitating concise yet accurate representations.

Overall, despite training the DNN exclusively on discrete AIF samples at defined time points, all DLIF configurations yield continuous, closed-form solutions capable of being evaluated at any arbitrary time. Although the number of active basis functions differs among configurations, each effectively captures the general shape and dynamics of the ground-truth AIF with commendable accuracy.

\subsection{Ablation and component analysis}
Next, we analyze the key elements that contribute to the improved quantitative performance of the proposed DLIF framework. To evaluate this, we used Pearson's correlation coefficient ($\pmb{r}$) to quantify the alignment of the DLIF shapes with the ground-truth AIFs. Additionally, we introduce a novel metric, Intersection-over-Union (IoU), which measures the overlap between the areas under the estimated and true input function curves. Unlike metrics focused on matching peak values, IoU emphasizes the integral areas, which are critical for downstream analyses such as Logan graphical analysis~\citep{logan1990graphical}. Details of these metrics are provided in the Methods section.

To refine the evaluation, we partition the input functions into two temporal segments: before and after 30 minutes, corresponding to the peak and stable phases of the input functions. Figure \ref{fig:ablation} presents the results of the component analysis using five-fold cross-validation on the validation datasets. The first plot in Fig. \ref{fig:ablation} depicts $\pmb{r}$ values, representing the shape correspondence between the DLIFs and the actual AIFs. Although direct estimation of input functions (labeled ``\textit{Direct Est.}'') achieves an average $\pmb{r}$ of 0.958, DLIF configurations using basis functions, either exponential or Gaussian, consistently outperform this approach. Specifically, the ``\textit{Exp.}'', ``\textit{Gaussian}'', ``\textit{Gaussian+Peak}'', and ``\textit{Gaussian+Peak+Sparse}'' configurations attain average $\pmb{r}$ values of 0.965, 0.964, 0.961, and 0.964, respectively. The second plot evaluates IoU, capturing the agreement in amplitudes and the total areas under the curves. Here, the ``\textit{Exp.}'', ``\textit{Gaussian}'', ``\textit{Gaussian+Peak}'', and ``\textit{Gaussian+Peak+Sparse}'' configurations achieve mean IoU scores of 0.787, 0.777, 0.781, and 0.780, respectively, slightly outperforming ``\textit{Direct Est.}'' (0.776).

Performance in the different temporal phases reveals further insight. During the peak phase (<30 minutes), the top-performing configurations, ``\textit{Exp.}'', ``\textit{Gaussian+Peak}'', and ``\textit{Gaussian+Peak+Sparse}'', record average IoU scores of 0.786, 0.781, and 0.779, respectively, compared to 0.776 for ``\textit{Direct Est.}''. During the tail phase (>30 minutes), these configurations achieve average IoU scores of 0.808, 0.803, and 0.799, respectively, compared to 0.780 for ``\textit{Direct Est.}''. Notably, the ``\textit{Gaussian}'' configuration slightly underperforms in the peak phase, with an average IoU of 0.776. 

Furthermore, the scatter plots demonstrate tighter distributions with shorter tails for ``\textit{Exp.}'', ``\textit{Gaussian+Peak}'', and ``\textit{Gaussian+Peak+Sparse}'' compared to ``\textit{Direct Est.}'', indicating more consistent performance. Based on these findings, we selected the top three configurations---``\textit{Exp.}'', ``\textit{Gaussian+Peak}'', and ``\textit{Gaussian+Peak+Sparse}''---for further quantitative and qualitative comparisons against baseline methods, as detailed in the subsequent section.

\begin{table*}
\centering
\fontsize{5}{7}\selectfont
\caption{Quantitative analysis of different methods on test sets not exposed during training. The
results are based on five-fold cross-validation. Note that the time intervals ``30-'' and ``30+'' refer to the periods of 0-30 minutes and 60-90 minutes, respectively. The top-performing scores are highlighted in \textbf{bold}. The symbol ``*'' denotes a statistically significant improvement, with a $p$-value < 0.05, as determined by a Wilcoxon signed-rank test with Bonferroni adjustment for multiple comparisons.}\label{tab:quant_res}
\begin{tabular}{ l|c l l l l l |c l l l l l} 
 \Xhline{1pt}
 Geno Types&\multicolumn{6}{c|}{HAB}&\multicolumn{6}{c}{MAB}\\
 \hline
   Methods  & $\pmb{r}$$\uparrow$ & IoU$\uparrow$ (30-) & IoU$\uparrow$ (30+) & RMSE$\downarrow$ (30-) & RMSE$\downarrow$ (30+) & Peak Bias$\downarrow$& $\pmb{r}$$\uparrow$ & IoU$\uparrow$ (30-) & IoU$\uparrow$ (30+) & RMSE$\downarrow$ (30-) & RMSE$\downarrow$ (30+) & Peak Bias$\downarrow$\\
 \Xhline{1pt}
 \rowcolor{Gray}
 EPICA & 0.50$\pm$0.29 & 0.39$\pm$0.16 & 0.32$\pm$0.26 & 2.34$\pm$1.22 & 0.23$\pm$0.23 & - & 0.52$\pm$0.27 & 0.36$\pm$0.13 & 0.34$\pm$0.28 & 3.48$\pm$1.14 & 0.25$\pm$0.12 & -\\
 
 ResNet50 & 0.90$\pm$0.11 & 0.62$\pm$0.15 & 0.55$\pm$0.14 & 1.72$\pm$1.10 & 0.14$\pm$0.10 & 0.52$\pm$0.52 & 0.90$\pm$0.09 & 0.57$\pm$0.17 & 0.56$\pm$0.15 & 2.36$\pm$1.15 & 0.14$\pm$0.07 & 0.45$\pm$0.20\\
 \rowcolor{Gray}
 ConvNeXt & 0.93$\pm$0.09 & 0.70$\pm$0.13 & 0.74$\pm$0.12 & 1.25$\pm$0.66 & 0.07$\pm$0.03 & 0.37$\pm$0.31 & \textbf{0.94$\pm$0.05} & \textbf{0.73$\pm$0.10} & 0.73$\pm$0.11 & 1.66$\pm$0.85 & 0.09$\pm$0.05 & 0.28$\pm$0.16\\
 
 DLIF-\textit{Exp.} & \textbf{0.94$\pm$0.08} & \textbf{0.72$\pm$0.13} & 0.73$\pm$0.14 & \textbf{1.13$\pm$0.66}* & 0.07$\pm$0.04 & 0.33$\pm$0.26 & 0.94$\pm$0.07 & 0.73$\pm$0.13 & 0.74$\pm$0.13 & \textbf{1.65$\pm$1.07} & 0.08$\pm$0.05 & 0.27$\pm$0.17\\
 \rowcolor{Gray}
 DLIF-\textit{Gaussian+Peak} & \textbf{0.94$\pm$0.08} & 0.71$\pm$0.13 & \textbf{0.76$\pm$0.14}* & 1.19$\pm$0.75 & 0.06$\pm$0.05 & 0.35$\pm$0.34 & 0.94$\pm$0.07 & 0.72$\pm$0.11 & \textbf{0.77$\pm$0.13}* & 1.70$\pm$0.91 & \textbf{0.07$\pm$0.05}* & 0.27$\pm$0.17\\
 
 DLIF-\textit{Gaussian+Peak+Sparse} & 0.93$\pm$0.09 & \textbf{0.72$\pm$0.13} & 0.75$\pm$0.13 & 1.17$\pm$0.74 & \textbf{0.06$\pm$0.04}* & \textbf{0.31$\pm$0.31}* & 0.94$\pm$0.07 & 0.72$\pm$0.13 & 0.77$\pm$0.15 & 1.67$\pm$1.01 & \textbf{0.07$\pm$0.05}* & \textbf{0.26$\pm$0.17} \\
 
 \Xhline{1pt}
\end{tabular}
\vspace{-3mm}
\end{table*}

\subsection{Qualitative and quantitative analysis}
Our study contrasts the proposed DLIF framework with the established traditional EPICA method~\citep{naganawa2005extraction, naganawa2005omission}. This method relies on decomposing uptake values categorized into different regions through independent component analysis. Originally, EPICA required invasive arterial sampling to obtain the peak value for scaling the estimated IDIF. However, for the purposes of our evaluation, we use the peak value from the actual ground-truth AIF as the scaling factor. Furthermore, given that DLIF represents a novel approach by integrating deep learning for input function estimation, it stands without direct predecessors employing a similar methodology. As a result, we benchmark the DLIF framework against two prominent DNNs in image classification, ResNet50~\citep{he2016deep} and ConvNeXt~\citep{liu2022convnet}. These models are adapted to produce discrete input function values at specified time points. They serve as strong comparatives to the ViT architecture within the DLIF framework, providing a comprehensive comparison within the deep learning domain.

For the quantitative evaluation of these methods, we extend our set of metrics beyond $\pmb{r}$ and IoU from the component analysis to include Root Mean Squared Error (RMSE) for assessing overall amplitude accuracy between the input functions, and percent peak bias for evaluating peak value correspondence. Given that EPICA uses the ground truth peak value for scaling, we exclude it from the peak bias assessment. Additionally, we analyze both the IoU and RMSE metrics across two distinct time intervals, before and after 30 minutes, to accurately capture the performance of the various methods during the initial peak phase and the tail of the input functions.

Fig. \ref{fig:quali_res} shows the qualitative results of the DLIFs under the three top performing configurations, ``\textit{Exp.}'', ``\textit{Gaussian+Peak}'', and ``\textit{Gaussian+Peak+Sparse}'', alongside the input functions estimated by EPICA, ResNet50, and ConvNeXt for four different subjects. The first two subjects are characterized by HAB genotypes, while the last two are MAB genotypes. The EPICA method noticeably falls short, failing to closely match the true AIF shapes. This is particularly evident in the zoom-in views, where EPICA often inaccurately identifies peak time points, resulting in jagged and inferior amplitude alignment in the input functions' tails. This limitation is likely due to its reliance on an unsupervised strategy that estimates AIFs by decomposing voxel values across the dynamic PET scan sequence using independent component analysis, which may be affected by noise presented in the PET images. In contrast, the DNN-based supervised learning approaches show significantly better adherence to the ground truth AIF shapes. Within this group, ResNet50 exhibits the least effectiveness in capturing both the shape and amplitude of the AIFs. ConvNeXt, while adept at pinpointing peak values and time points, tends to produce results that are not as smooth, with less accurate amplitude matching post-peak. Upon closer examination, highlighted in the second and third rows of Fig. \ref{fig:quali_res}, it is evident that the proposed DLIF framework generates AIFs with a smoother consistency and a closer match to the ground truth across both patient genotypes compared to other methods, which show greater fluctuations in their estimated input functions.

\begin{figure*}[t]
\begin{center}
\includegraphics[width=0.99\textwidth]{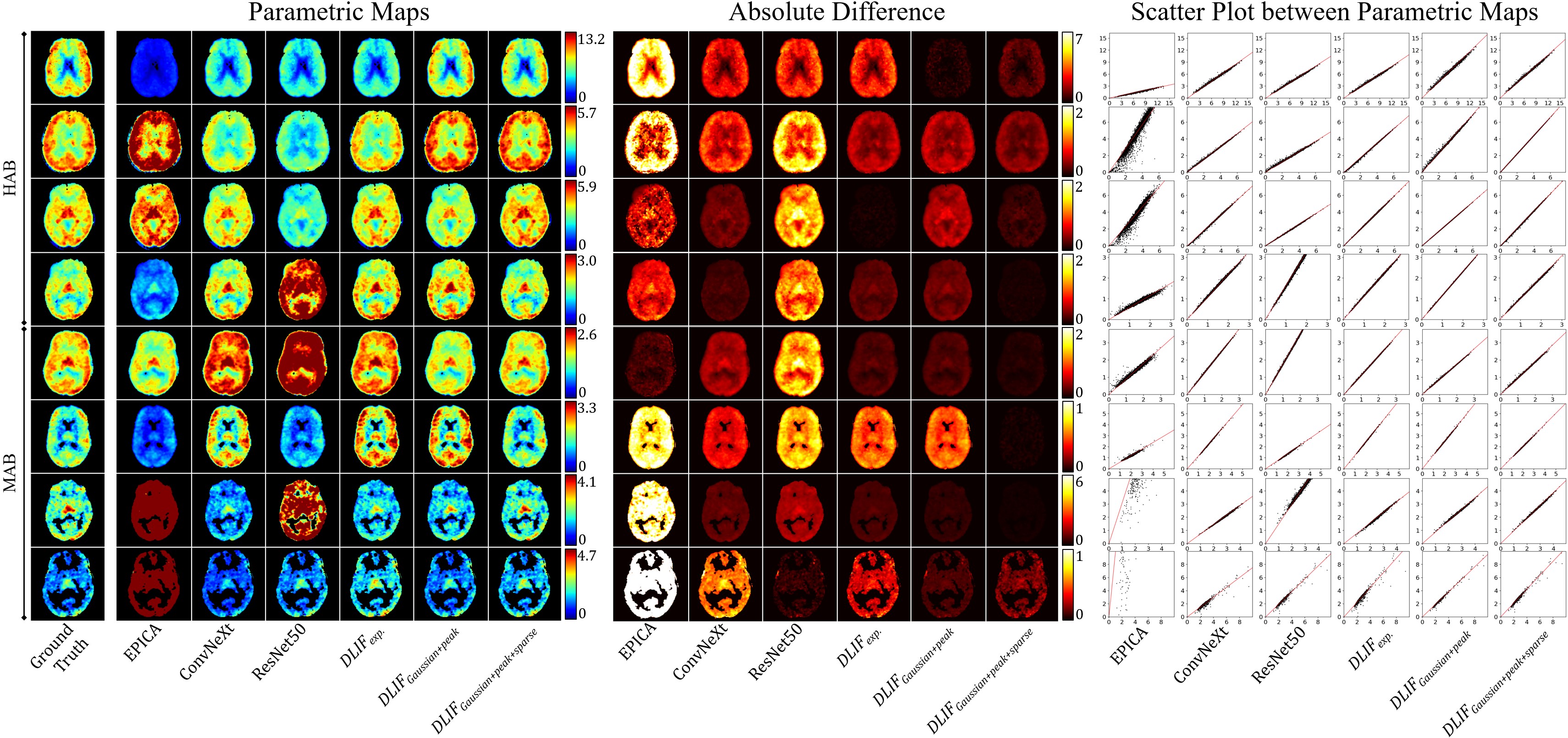}
\end{center}
   \caption{This figure illustrates the parametric maps (left) reconstructed based on AIFs and their percentage differences relative to the ground truth (right). Each row represents a different subject, and each column represents a different AIF estimation method. The subjects are grouped according to their genotypes: C/C or C/T. \textcolor{black}{For visualization, subject-specific color scales are used to account for the large inter-subject variability in intensity ranges.}}
\label{fig:para_results}
\end{figure*}

Table \ref{tab:quant_res} showcases the quantitative analysis comparing various methods, with the top performances emphasized in bold. Statistical significance tests, specifically the Wilcoxon signed-rank tests, were conducted between the top-performing non-DLIF model, ConvNeXt, and the top DLIF models to assess significance, with Bonferroni correction applied to results with the identical scores. The quantitative results for EPICA, which correlate with its qualitative evaluations, indicate it as the least effective method. It shows mean $\pmb{r}$ values of approximately 0.5 and suboptimal mean IoU values of about 0.35, demonstrating poor correspondence in both shape and integral matching of the input functions. In contrast, the deep learning-based methods surpass EPICA across all metrics, with the proposed DLIF models showing exceptional performance, particularly achieving mean $\pmb{r}$ values of 0.94 for both HAB and MAB patients, which indicates strong shape agreement with the ground truth AIFs from arterial blood sampling. Look at the other metrics, the ``\textit{Gaussian+Peak}'' configuration of the proposed DLIF demonstrated statistically significant improvements in IoU and RMSE for the tail of the AIFs (i.e., IoU (30+)), with mean IoU values of 0.76 and 0.77, and mean RMSE values of 0.06 and 0.07, respectively, for HAB and MAB genotypes. The ``\textit{Gaussian+Peak+Sparse}'' configuration yielded the best peak matching, showing the least peak bias with values of 0.31 and 0.26 for HAB and MAB genotypes respectively. In the peak regions (i.e., <30 mins or 30-), all deep learning-based methods performed reasonably well, with the proposed DLIFs achieving the highest mean IoU and RMSE values, surpassing ResNet50 significantly and being slight better than the results of ConvNeXt.

\begin{figure*}[t]
\begin{center}
\includegraphics[width=0.99\textwidth]{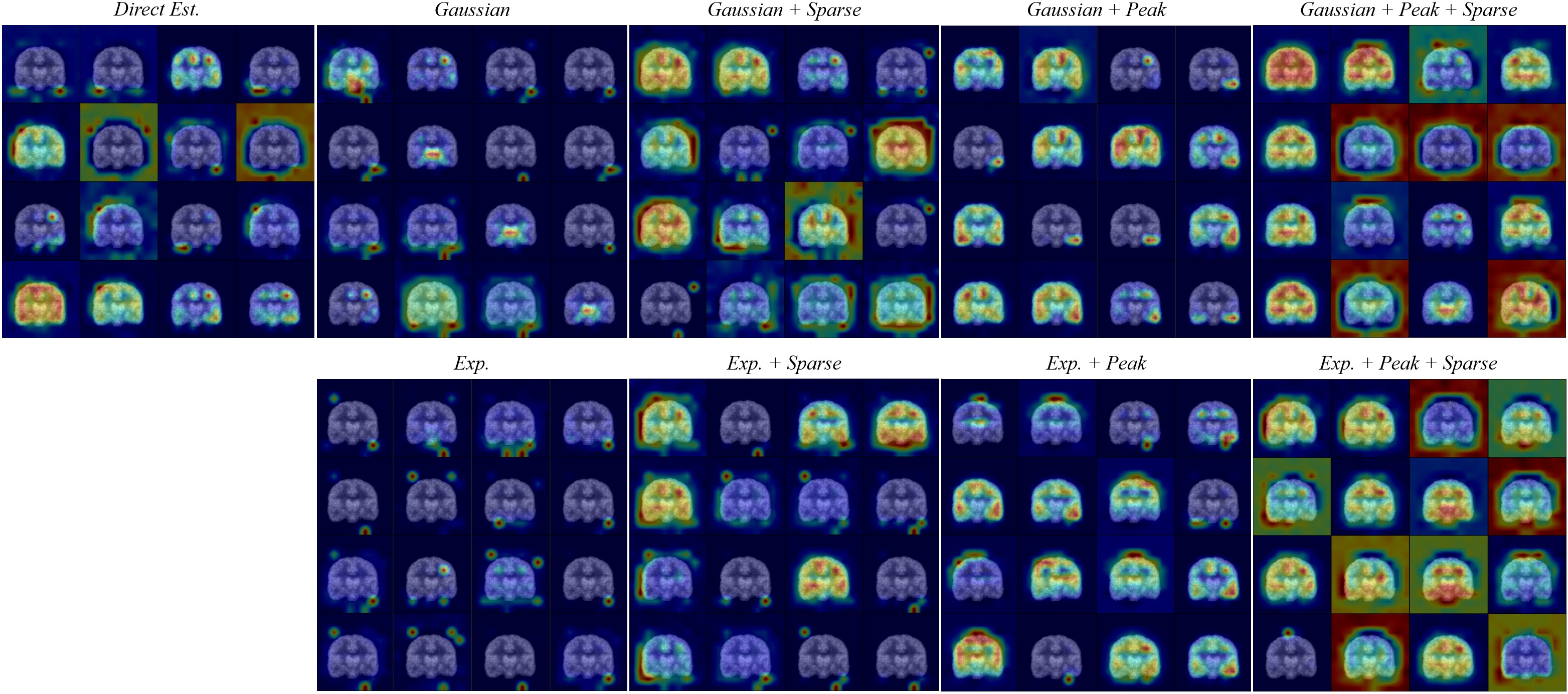}
\end{center}
   \caption{Visualization of attention maps provided by the ViT of the proposed DLIF framework, showcasing nine distinct configurations. These range from direct estimation of the AIFs (``\textit{Direct Est.}'') to implementations using basis functions, both with and without the integration of peak estimation and sparsity constraints.}
\label{fig:attention}
\end{figure*}

\subsection{Logan graphical analysis}
We also undertook secondary, indirect comparisons using the parametric maps generated from the estimated input functions. These brain distribution volume maps ($V_T$) were derived using Logan graphical analysis, as detailed in the Methods section~\citep{logan1990graphical}. Figure~\ref{fig:para_results} provides qualitative comparisons of these parametric maps for eight patients spanning two genotypes: the upper four rows illustrate HAB genotypes, and the lower four rows show MAB genotypes. The left panel displays the parametric maps, the middle panel highlights absolute differences between estimated and ground truth values, and the right panel features scatter plots comparing voxel-wise estimations to ground truth values.

Inspection of these parametric maps reveals that EPICA-generated results exhibited substantial deviations from the ground truth, with pronounced over- and underestimations across subjects. In contrast, all deep learning-based methods consistently produced input functions closer to the ground truth. Among these, ResNet50 was notably less effective, aligning with its previously documented limitations in AIF estimation. The DLIF approach proposed here demonstrated the most accurate visual representations within parametric maps. These qualitative observations are quantitatively supported by the absolute difference images (middle panel), which clearly illustrate that EPICA produced large voxel value discrepancies, whereas DLIF approaches displayed significantly smaller differences, outperforming other deep learning methods. Scatter plots (right panel), based on 10,000 randomly selected voxels within brain regions, further confirm these results, showing strong correlations for all deep learning methods with ground truth values. In particular, DLIF-generated results yielded regression slopes approaching unity, underscoring their high fidelity to ground truth.

Table~\ref{tab:quant_res_VT} summarizes quantitative assessments of parametric maps estimated from the various AIF methods. Remarkably, DLIF-\textit{Gaussian+Peak} achieved superior performance, presenting the lowest mean MAE (0.66) and RMSE (0.68). ConvNeXt closely followed, with similarly robust performance (mean MAE of 0.66, mean RMSE of 0.69). Conversely, ResNet50 and EPICA delivered notably poorer results, with EPICA, despite requiring invasive arterial sampling at peak time points, demonstrating the weakest performance overall.

\begin{table}
    \centering
    \fontsize{9}{11}\selectfont
    \caption{Quantitative analysis of parametric maps estimated based on AIFs generated by different methods.}\label{tab:quant_res_VT}
    \begin{tabular}{ l|c c } 
    \Xhline{1pt}
    Methods & MAE$\downarrow$ & RMSE$\downarrow$\\
    \Xhline{1pt}
     \rowcolor{Gray}
     EPICA & 3.51$\pm$6.32 & 3.65$\pm$6.55 \\
     
     ResNet50 & 1.24$\pm$1.00 & 1.27$\pm$1.02\\
     \rowcolor{Gray}
     ConvNeXt & 0.67$\pm$0.64 & 0.69$\pm$0.66\\
    
     DLIF-\textit{Exp.}  & 0.75$\pm$0.73 & 0.77$\pm$0.75\\
     \rowcolor{Gray}
     DLIF-\textit{Gaussian+Peak}  & \textbf{0.66$\pm$0.69} & \textbf{0.68$\pm$0.70} \\
     
     DLIF-\textit{Gaussian+Peak+Sparse} & 0.71$\pm$0.80 & 0.73$\pm$0.81\\
    \Xhline{1pt}
    \end{tabular}
\end{table}

\section{Discussion}
\label{sec:discussion}
\subsection{\textcolor{black}{Performance of DLIF Configurations and Evaluation Metrics}}
We performed extensive experiments to optimize the selection and configuration of basis functions within the DLIF framework and compared its performance against existing IDIF methods and alternative deep learning approaches that directly estimate discrete AIF values. Analysis of the initial two plots in Fig.~\ref{fig:ablation} showed that all tested DLIF configurations strongly correlated with the ground-truth AIF shapes, yielding average correlation coefficients ($\pmb{r}$) exceeding 0.9. Such high correlation demonstrates that the DLIF effectively captures metabolite-corrected AIF shapes and underscores its suitability in clinical scenarios where a single arterial sample could be available for scaling, mirroring conventional clinical practices.

To further compare the estimated and ground truth AIFs, we introduced a novel evaluation metric based on the ratio of the integrated area of overlap, or IoU. This metric emphasizes the integral areas under the curves, which is physiologically important for determining total distribution volume, rather than focusing solely on matching peak values at specific time points. Based on the $\pmb{r}$ and IoU metrics, we identified three particularly effective configurations of the DLIF on the validation dataset: ``\textit{Exp.}'', ``\textit{Gaussian+Peak}'', and ``\textit{Gaussian+Peak+Sparse}''. We compared these configurations against various baseline methods on the test set. Qualitative results in Fig.~\ref{fig:quali_res} confirmed the capability of DLIF to match the peak of ground truth while producing a smooth tail that closely matches the original data. Other methods resulted in a less smooth tail, with the traditional method EPICA performing the lowest, as its estimated input function struggled to match the ground truth even though it required the ground truth peak value as a scaling factor. Quantitative results shown in Table~\ref{tab:quant_res} further validate our findings, with DLIF configurations significantly outperforming others. When comparing the parametric maps estimated using the AIFs predicted by various methods, all DLIF configurations showed a good match with the ground truth, as indicated by the absolute difference images and the scatter plots shown in Fig.~\ref{fig:para_results}. The ``\textit{Gaussian+Peak}'' configuration demonstrated the least quantitative error, as detailed in Table~\ref{tab:quant_res_VT}.

\subsection{\textcolor{black}{Genotype-Specific Analysis}}
We also investigated whether the model outcomes would differ according to the genotypes of the subjects (i.e., HAB and MAB). Table~\ref{tab:p_values_geno} presents the $p$-values of the Mann-Whitney U test, comparing the quantitative scores in two distinct genotype groups. The results indicate that all models showed no statistically significant differences in $\pmb{r}$ values, IoU metrics, or peak bias, suggesting that the models produce similar AIF shapes, peak values, and integrals regardless of the genotype differences. However, there was a statistically significant difference in RMSE during the first 30 minutes across all models, indicating that the average magnitude of the prediction errors significantly varies between the two groups during the peak phase. This discrepancy is likely due to inherent differences in peak AIF values and the uptake of pharmaceuticals in brain tissues between genotypes. Statistical analysis confirmed significant differences in the mean and peak AIF values between the two genotypes, with $p$-values $\ll 0.001$. Conversely, the proposed DLIFs reported non-significant RMSE scores in predictions for the tail phase (i.e., longer than 30 minutes). In contrast, both ConvNeXt and EPICA exhibited significant differences in RMSE scores between the two genotype groups during this phase. The consistency in RMSE observed with the DLIFs beyond 30 minutes likely benefits from the inclusion of basis functions, demonstrating robust performance in the later stages of AIF analysis despite variations in genotypes.

\begin{table*}
\centering
\fontsize{9}{11}\selectfont
\caption{$p$-values derived from the Mann-Whitney U test, adjusted for multiple comparisons using a Bonferroni correction factor of 6, comparing quantitative scores between two genotypes. Values in \textit{italic} denote statistical significance (i.e., $p<0.05$).}\label{tab:p_values_geno}
\begin{tabular}{ l|c c c c c c} 
 \Xhline{1pt}
 &\multicolumn{6}{c}{$p$-values from the Mann-Whitney U Test Comparing Genotypes}\\
 \hline
 Methods   & $\pmb{r}$ & IoU (30-) & IoU (30+) & RMSE (30-) & RMSE (30+) & Peak Bias\\
 \Xhline{1pt}
 \rowcolor{Gray}
 EPICA & 2.559 & 1.185 & 2.678 & \textit{0.000} & \textit{0.037} & - \\
 ResNet50 & 1.445 & 0.458 & 1.787 & \textit{0.003} & 0.701 & 1.102 \\
 \rowcolor{Gray}
 ConvNeXt & 2.885 & 0.883 & 1.417 & \textit{0.014} & \textit{0.032} & 0.677\\
 
 DLIF-\textit{Exp.} & 2.643 & 1.271 & 2.475 & \textit{0.009} & 0.289 & 1.135\\
 \rowcolor{Gray}
 DLIF-\textit{Gaussian+Peak} & 2.428 & 2.082 & 2.511 & \textit{0.001} & 0.277 & 1.168\\
 
 DLIF-\textit{Gaussian+Peak+Sparse} & 2.872 & 2.933 & 0.429 & \textit{0.013} & 1.008 & 2.559 \\
 
 \Xhline{1pt}
\end{tabular}
\vspace{-3mm}
\end{table*}

\subsection{\textcolor{black}{Architectural Flexibility of DLIF}}
The quantitative scores presented in Tables~\ref{tab:quant_res} and \ref{tab:quant_res_VT} show that ConvNeXt performs comparable to the proposed method. This outcome aligns with the foundational design objectives of ConvNeXt, which was conceived as a robust alternative to vision Transformers. It is imperative to note that the proposed DLIF framework is designed to be agnostic of any specific network architecture, thus facilitating the easy incorporation of future sophisticated architectures to potentially enhance performance measures. As such, ConvNeXt could be effectively used as the backbone for the DLIF framework. However, the decision to employ ViT in this research is based on its ability to visualize attention mechanisms, providing an explainable model for the network predictions, which is essential for validation and interpretation in clinical contexts. 

\begin{table*}
\centering
\fontsize{7.5}{9.5}\selectfont
\caption{\textcolor{black}{Quantitative results of different methods from the generalizability and transfer learning studies. The results are based on five-fold cross-validation. Note that the time intervals ``30-'' and ``30+'' refer to the periods of 0-30 minutes and 60-90 minutes, respectively. The top-performing scores are highlighted in \textbf{bold}. The symbol ``*'' denotes a statistically significant improvement between without and with transfer learning (i.e., fine-tuning), with a $p$-value < 0.05, as determined by a Wilcoxon signed-rank test.}}
\label{tab:adapt}
\begin{tabular}{ l|c c c |c c c } 
 \Xhline{1pt}
 \multicolumn{7}{c}{[\textsuperscript{11}C]DPA-713 PET data acquired on the Siemens Biograph mCT scanner (New Scanner)} \\
 \hline
 & \multicolumn{3}{c|}{Without fine-tuning (zero-shot)} & \multicolumn{3}{c}{With fine-tuning} \\
 \hline
   Metrics  & DLIF-\textit{Exp.} & DLIF-\textit{Gaussian+Peak} & DLIF-\textit{Gaussian+Peak+Sparse} & DLIF-\textit{Exp.} & DLIF-\textit{Gaussian+Peak} & DLIF-\textit{Gaussian+Peak+Sparse} \\
 \Xhline{1pt}
 \rowcolor{Gray}
 $\pmb{r}$$\uparrow$              & 0.95$\pm$0.05 & 0.95$\pm$0.06 & 0.95$\pm$0.06 & 0.96$\pm$0.05 & 0.96$\pm$0.08* & \textbf{0.97$\pm$0.05}* \\
 
 IoU$\uparrow$ (30-)              & 0.74$\pm$0.11 & 0.73$\pm$0.12 & 0.71$\pm$0.11 & 0.80$\pm$0.09* & 0.80$\pm$0.08* & \textbf{0.81$\pm$0.09}* \\
 \rowcolor{Gray}
 IoU$\uparrow$ (30+)              & 0.78$\pm$0.09 & 0.83$\pm$0.12 & 0.79$\pm$0.14 & 0.82$\pm$0.09 & \textbf{0.84$\pm$0.09} & 0.83$\pm$0.07 \\
 
 RMSE$\downarrow$ (30-)           & 1.82$\pm$1.05 & 1.89$\pm$1.12 & 1.98$\pm$1.18 & 1.39$\pm$0.90* & 1.39$\pm$0.93* & \textbf{1.35$\pm$1.01}* \\
 \rowcolor{Gray}
 RMSE$\downarrow$ (30+)           & 0.07$\pm$0.03 & 0.05$\pm$0.04 & 0.06$\pm$0.05 & 0.06$\pm$0.04 & 0.05$\pm$0.04 & \textbf{0.05$\pm$0.03} \\
 
 Peak Bias$\downarrow$            & 0.33$\pm$0.15 & 0.34$\pm$0.17 & 0.35$\pm$0.13 & 0.20$\pm$0.14* & 0.22$\pm$0.11* & \textbf{0.17$\pm$0.14}* \\
 \hline
 \hline
  \multicolumn{7}{c}{[\textsuperscript{11}C]CPPC PET data acquired on the Siemens Biograph mCT scanner (New tracer \& scanner)} \\
 \hline
 & \multicolumn{3}{c|}{Without fine-tuning (zero-shot)} & \multicolumn{3}{c}{With fine-tuning} \\
 \hline
   Metrics  & DLIF-\textit{Exp.} & DLIF-\textit{Gaussian+Peak} & DLIF-\textit{Gaussian+Peak+Sparse} & DLIF-\textit{Exp.} & DLIF-\textit{Gaussian+Peak} & DLIF-\textit{Gaussian+Peak+Sparse} \\
 \Xhline{1pt}
 \rowcolor{Gray}
 $\pmb{r}$$\uparrow$              & 0.75$\pm$0.23 & 0.68$\pm$0.18 & 0.83$\pm$0.10 & 0.90$\pm$0.16 & 0.91$\pm$0.17* & \textbf{0.93$\pm$0.12}* \\
 
 IoU$\uparrow$ (30-)              & 0.46$\pm$0.12 & 0.43$\pm$0.11 & 0.40$\pm$0.13 & 0.76$\pm$0.14* & 0.77$\pm$0.14* & \textbf{0.79$\pm$0.15}* \\
 \rowcolor{Gray}
 IoU$\uparrow$ (30+)              & 0.55$\pm$0.19 & 0.64$\pm$0.13 & 0.66$\pm$0.15 & 0.79$\pm$0.10* & \textbf{0.80$\pm$0.10}* & 0.73$\pm$0.11 \\
 
 RMSE$\downarrow$ (30-)           & 3.09$\pm$0.77 & 3.37$\pm$0.63 & 3.19$\pm$0.62 & 1.36$\pm$0.99* & 1.51$\pm$1.24* & \textbf{1.22$\pm$1.05}* \\
 \rowcolor{Gray}
 RMSE$\downarrow$ (30+)           & 0.10$\pm$0.08 & 0.06$\pm$0.03 & 0.07$\pm$0.06 & \textbf{0.03$\pm$0.01}* & \textbf{0.03$\pm$0.01}* & 0.04$\pm$0.02 \\
 
 Peak Bias$\downarrow$            & 0.73$\pm$0.15 & 0.80$\pm$0.14 & 0.79$\pm$0.14 & \textbf{0.27$\pm$0.37}* & 0.36$\pm$0.55 & 0.32$\pm$0.59* \\
 \Xhline{1pt}
\end{tabular}
\vspace{-3mm}
\end{table*}

\subsection{\textcolor{black}{Interpretability through Attention Maps}}
DL models are often described as ``black-box'' models due to their limited ability to explain how specific estimates are derived. However, a notable advantage of using ViT within the DLIF framework is its inherent capability to generate attention maps, elucidating the regions of focus during prediction. Figure~\ref{fig:attention} illustrates attention map visualizations from the final self-attention layer of the ViT for sixteen randomly selected datasets across various DLIF configurations. Given that DLIF aims to directly estimate metabolite-corrected AIFs from dynamic PET images reflecting radiotracer uptake in the brain, it is expected that the ViT predominantly attends to regions of high uptake (higher intensity voxels). In the ``\textit{Direct Est.}'' configuration, while the ViT's attention is generally centered on the brain tissue, it occasionally diverts to less pertinent background regions. The introduction of basis functions without peak estimation or sparsity enforcement, seen in the ``\textit{Gaussian}'' and ``\textit{Exp.}'' configurations, led to attention mechanisms occasionally fixating on confined background regions, making interpretation non-intuitive. Imposing sparsity constraints slightly refined the attention mechanisms' focus within the brain, although some attention still lingered on the background. In contrast, integrating peak estimation markedly improved focus on brain regions, with attention spreading to areas of high uptake, especially evident in the ``\textit{Gaussian+Peak}'' and ``\textit{Exp.+Peak}'' configurations. However, adding a sparsity constraint redirected the attention mechanisms towards the background again, possibly because enforcing sparsity prompted the ViT to leverage small values in these regions to assign smaller weights to certain basis functions.

\subsection{\textcolor{black}{Generalizability of the DLIF Framework}}
\textcolor{black}{We evaluated the generalizability of the proposed DLIF framework by testing it on two new tasks designed to probe its adaptability and potential for transfer learning. First, to assess robustness to hardware variability, we applied DLIF to 21 [\textsuperscript{11}C]DPA-713-TSPO-PET dynamic scans acquired on a newer generation scanner (Siemens Biograph mCT, a time-of-flight whole body system). Second, to examine adaptability across tracers, we applied DLIF to 10 [\textsuperscript{11}C]CPPC-CSF1R-PET dynamic scans acquired with [\textsuperscript{11}C]CPPC on the Siemens Biograph mCT system~\citep{coughlin2022first,horti2019pet,mills2025exploring}. [\textsuperscript{11}C]CPPC binds to a different target, CSF1R, and exhibits pharmacokinetic properties very distinct from [\textsuperscript{11}C]DPA-713, which binds to TSPO. Compared to [\textsuperscript{11}C]DPA-713, [\textsuperscript{11}C]CPPC exhibits a slower washout and lower peak in the AIF, representing behavior not encountered during the original DLIF training. All PET images were normalized to SUVs and the datasets included associated metabolite-corrected AIFs.
}

\begin{figure*}[thp]
\begin{center}
\includegraphics[width=0.99\textwidth]{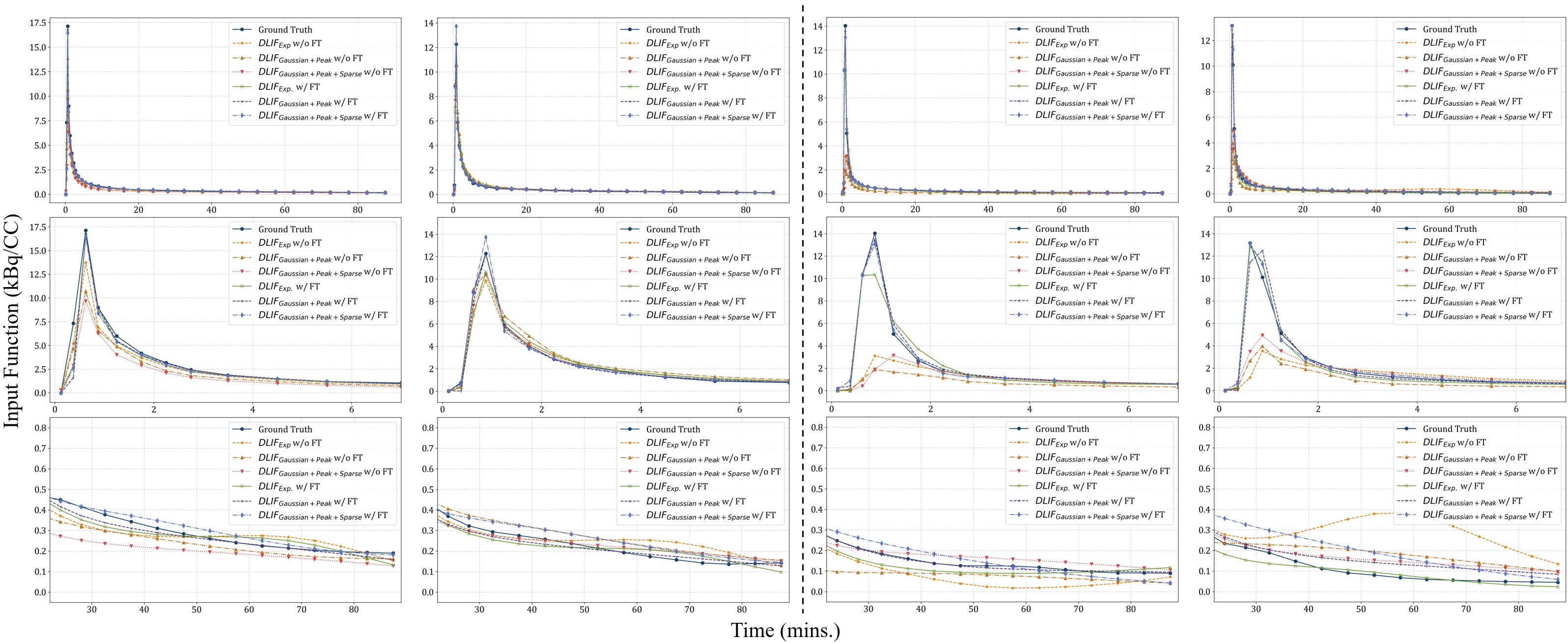}
\end{center}
   \caption{\textcolor{black}{Qualitative results for the generalizability and transfer learning tasks. The left panel shows AIFs and DLIF estimates from [\textsuperscript{11}C]DPA-713-TSPO-PET data acquired on a newer-generation scanner (Siemens Biograph mCT), while the right panel presents results from [\textsuperscript{11}C]CPPC-CSF1R-PET data acquired on a scanner of the same model type. Each column corresponds to a unique subject. The first row presents the full estimated input functions, and the second and third rows show zoomed-in views of the 0–7 minute and 22–90 minute intervals, respectively. ``w/o FT'' and ``w/ FT'' in the legend indicate results obtained without and with fine-tuning, respectively.}}
\label{fig:quali_ft}
\end{figure*}

\textcolor{black}{For both tasks, we first evaluated the zero-shot capability of DLIF framework by directly applying the model trained on [\textsuperscript{11}C]DPA-713 data from HRRT scanners. We then fine-tuned the model by allowing only the estimation heads and the final Transformer block to remain trainable, while freezing all other parameters. The model was fine-tuned for 100 epochs with a learning rate of 1e-4. This transfer learning strategy reduces the number of trainable parameters, enabling effective adaptation to new domains under limited data availability while mitigating overfitting. We performed five-fold cross-validation for both tasks, with each fold using 8:2 split for training and evaluation.
}

\subsubsection{\textcolor{black}{New scanner}}
\label{sec:new_scanner}
\textcolor{black}{The left panel of Fig.~\ref{fig:quali_ft} shows two examples of DLIF estimates compared with the ground-truth AIFs. Despite the change in scanners, the DLIF outputs without fine-tuning closely matched the reference AIFs, though with slightly lower peak values. This observation is corroborated by the quantitative results in the upper panel of Table~\ref{tab:adapt}, where zero-shot application of the DLIF framework to the new scanner data achieved performance comparable to that reported in Table~\ref{tab:quant_res}, which was obtained from test images drawn from the same distribution as the training dataset. Specifically, the correlation coefficient ($\pmb{r}$) remained around 0.95, IoU values exceeded 0.7, and both RMSE and peak bias were relatively small across all three DLIF variants. These findings indicate that the proposed DLIF framework achieves strong zero-shot performance for [\textsuperscript{11}C]DPA-713 scans acquired on different scanners, suggesting that it captures prior knowledge of [\textsuperscript{11}C]DPA-713 AIFs in a manner largely invariant to scanner differences. Further investigation is needed to determine whether this robustness extends to additional scanner types and acquisition protocols. }

\textcolor{black}{Although the zero-shot performance was already strong, additional improvements were achieved through fine-tuning. Qualitative results in Fig.~\ref{fig:quali_ft} demonstrate a closer match between the DLIF estimates and the ground-truth AIFs after fine-tuning. Quantitative scores in Table~\ref{tab:adapt} also showed statistically significant gains, particularly in shape and peak matching, with the best performance achieved by the ``\textit{Gaussian+Peak+Sparse}'' variant. Improvements were most pronounced in peak-related metrics: IoU and RMSE within the first 30 minutes, as well as peak bias, all showed significant improvement relative to their zero-shot counterparts, whereas performance during the later 60–90 minute period remained comparable.  
}

\subsubsection{\textcolor{black}{New tracer \& scanner}}
\textcolor{black}{The right panel of Fig.~\ref{fig:quali_ft} illustrates results for [\textsuperscript{11}C]CPPC. In this case, the pretrained DLIF model failed to produce accurate predictions, with both the peak amplitude and overall AIF shape deviating substantially from the ground truth. Quantitative results in the lower panel of Table~\ref{tab:adapt} confirm this observation: without fine-tuning, the correlation coefficient ($\pmb{r}$) remained in the suboptimal range of 0.7–0.8, IoU values were consistently below 0.7, and both RMSE and peak bias were relatively high. These findings support our hypothesis that a model trained exclusively on [\textsuperscript{11}C]DPA-713 data develops prior knowledge specific to [\textsuperscript{11}C]DPA-713 AIF characteristics, which enables robustness to scanner variability but limits generalization to tracers unseen from training with distinct pharmacokinetics such as [\textsuperscript{11}C]CPPC.}

\textcolor{black}{
The key question is whether the DLIF framework trained on [\textsuperscript{11}C]DPA-713 can be effectively transferred to [\textsuperscript{11}C]CPPC, which was also acquired on a different scanner (mCT instead of HRRT). The results demonstrate that it can: with only eight subjects used for fine-tuning in each fold, performance \textcolor{black}{improved substantially. This improvement is evident in both the qualitative results (Fig.~\ref{fig:quali_ft}) and the quantitative results (Table~\ref{tab:adapt}), where the scores became comparable to those obtained on [\textsuperscript{11}C]DPA-713 data (Table~\ref{tab:quant_res}). Consistently, most metrics also showed statistically significant improvements relative to their zero-shot counterparts.}
}

\textcolor{black}{In this work, the metabolite-corrected AIFs were used as ground truth for training; therefore, the DLIF framework directly predicts metabolite-corrected AIFs without explicitly modeling metabolite kinetics. For tracers that do not produce metabolites, require whole-blood rather than metabolite-corrected input functions, or exhibit tracer-specific metabolite behavior such as metabolites crossing the blood–brain barrier, additional training or transfer learning using the appropriate input functions would be required. Despite these tracer-specific considerations, the underlying DLIF framework is general and can be adapted beyond the current scope. Although this study focused on TSPO-PET imaging of the brain, the framework can, with fine-tuning on small tracer-specific datasets or retraining with new data, be extended to other tracers, anatomical regions, and imaging protocols. Overall, these findings highlight the potential of DLIF to provide accurate and fully non-invasive estimates of arterial input functions across diverse PET imaging applications.}

\subsection{\textcolor{black}{Compartment Analysis using DLIF}}
\textcolor{black}{
Given the close agreement between the DLIF- and AIF-derived curves, we further investigated whether DLIF could be extended beyond Logan graphical analysis to support full compartmental modeling. As a proof of concept, we analyzed a representative [\textsuperscript{11}C]DPA-713 scan acquired on the Siemens Biograph mCT scanner (as described in Section~\ref{sec:new_scanner}) using the fine-tuned DLIF model for AIF estimation. A two-tissue compartment model was applied, and the total distribution volume ($V_T$) values across multiple brain regions were compared with those obtained using ground-truth AIFs. The results are presented in \ref{sec:comp_model} (Fig.~\ref{fig:two_fit}), which shows a correlation plot where the $x$-axis represents $V_T$ estimates derived from ground-truth AIFs and the $y$-axis represents those derived from DLIF, with each circle corresponding to a distinct brain region. The correlation between DLIF- and AIF-based $V_T$ estimates was close to unity (dashed line), indicating strong agreement. An example fit for the frontal cortex further demonstrates that the DLIF- and AIF-driven model fits closely overlap. This proof-of-concept analysis illustrates that DLIF-derived AIFs can reliably support compartmental kinetic modeling, highlighting their potential utility beyond simplified approaches such as Logan graphical analysis.}

\section{Conclusion}
\label{sec:conclusion}
In this study, we introduced DLIF, a deep learning-based framework designed to non-invasively estimate AIF, entirely eliminating the necessity of invasive arterial blood sampling. The DLIF framework incorporates prior physiological knowledge from existing literature by representing AIFs using continuous basis functions, such as Gaussian and exponential-sigmoid functions. By leveraging the flexibility of DNNs, DLIF accurately models complex temporal dynamics through an overcomplete representation of these basis functions, producing input functions as combinations of individually predicted parameters. Unlike traditional methods, including IDIF and PBIF, which yield a non-personalized estimation of AIF~\citep{Zanotti_Fregonara_2012, Boutin_2007} or often require a scaling factor from single or multiple time points invasive arterial or venous blood sampling to calculate AIFs for patients~\citep{zanotti2011image, Takikawa_1993, naganawa2005extraction, naganawa2005omission,naganawa2008robust,litton1997input,chen1998noninvasive}, DLIF completely avoids such invasive procedures. It offers fast and robust estimation of AIFs using only the subjects' dynamic PET sequences. This makes the DLIF framework a promising tool for broader adoption of dynamic PET imaging in routine clinical settings, replacing current semi-quantitative analysis methods. This advancement could substantially enhance diagnostic accuracy for neurological conditions such as Alzheimer's disease and Parkinson's disease, particularly during early-stage evaluations.

\section{Acknowledgments}
This work was supported by the Macks Family Foundation and the Michael J. Fox Foundation, as well as by grants from the National Institutes of Health (R01EB031023, P01CA272222, U01CA140204, R01NS100847, and P41EB024495). The content is solely the responsibility of the authors and does not necessarily represent the official views of the NIH. 
\onecolumn 
\appendix
\section{Vision Transformer and Self-Attention}
\label{sec:vit}
\textcolor{black}{
As depicted in Fig. \ref{fig:net_arch}, the downsampled dynamic PET image $I\in\mathbb{R}^{T\times\frac{H}{4}\times\frac{W}{4}\times\frac{L}{4}}$ is first divided into a series of $N$ non-overlapping patches $\{I_1, I_2, ..., I_N\}$, each of size $T\times P\times P\times P$, that is, $I_i\in\mathbb{R}^{T\times P\times P\times P}$. Subsequently, each patch is vectorized and linearly projected to create tokens with a $D$-dimensional embedding. This is formulated as:
\begin{equation}
    \hat{\mathbf{z}}_0 = \{I_1\mathbf{E}, I_2\mathbf{E},..., I_N\mathbf{E}\},\ \ \mathbf{E}\in\mathbb{R}^{TP^3\times D},
\end{equation}
so that $\hat{\mathbf{z}}$ assumes the shape of $N\times D$. Next, we append separate learnable embeddings/tokens ($I^\text{spatial}\in\mathbb{R}^{1\times D}$) to both the sequence of embedded patches and the temporal dimension, respectively. Consequently, $\hat{\mathbf{z}}_0$ is expanded to a size of $(N+1)\times D$. This token interacts with all other tokens across spatial dimensions, which resembles the traditional methods that compare voxels across varying spatial distances~\citep{wang2006model, naganawa2008robust}. In a later stage, a specifically designed estimation head will be attached to $I^\text{spatial}$ to generate the DLIF. Finally, before progressing to the Transformer encoder, a learnable positional embedding is added to the tokens, enabling them to preserve spatial information throughout the network. The process is described as:
\begin{equation}
    \mathbf{z}_0 = \hat{\mathbf{z}}_0+\mathbf{E}_{pos},\ \ \mathbf{E}_{pos}\in\mathbb{R}^{(N+1)\times D}.
\end{equation}
The resulting tokens are then processed through a Transformer encoder, comprising $J$ sequential base blocks. Each block contains two four components as shown in Fig. \ref{fig:net_arch}, and it is mathematically represented as:
\begin{equation}
    \begin{split}
        \hat{\mathbf{z}}_\ell=\text{MSA}(\text{LN}(\mathbf{z}_{\ell-1}))+\mathbf{z}_{\ell-1},\ \ \ \ &\ell=1...L,\\
        \mathbf{z}_\ell=\text{MLP}(\text{LN}(\hat{\mathbf{z}}_\ell))+\hat{\mathbf{z}}_\ell,\ \ \ \ &\ell=1...L,\\
    \end{split}
\end{equation}
where $\text{LN}(\cdot)$ indicates layer normalization~\citep{ba2016layer}. The operations $\text{MSA}(\cdot)$ and $\text{MLP}(\cdot)$ represent the multi-head self-attention mechanism and the multi-layer perceptron, respectively.}

\textcolor{black}{Here, we adopt the standard softmax-based self-attention operation~\citep{vaswani2017attention}, where the tokens $\mathbf{z}$ are first projected into three separate embeddings: a query $\mathbf{Q}$, a key $\mathbf{K}$, and a value $\mathbf{V}$ using a set of learnable parameters $\mathbf{U}_{qkv}$:
\begin{equation}
\begin{split}
    [\mathbf{Q}, \mathbf{K}, \mathbf{V}]=\mathbf{z}\mathbf{U}_{qkv}&,\ \ \ \ \mathbf{U}_{qkv}\in\mathbb{R}^{D\times3D_h},\\
    \mathbf{A} = \text{softmax}(\frac{\mathbf{Q}\mathbf{K}^T}{\sqrt{D_h}})&,\ \ \ \ \mathbf{A}\in\mathbb{R}^{(N+1)\times (N+1)},\\
    \text{SA}(\mathbf{z})=\mathbf{A}\mathbf{V}.&
\end{split}
\end{equation}}

\section{Results of Two-Tissue Compartment Model Analysis Using DLIF}
\label{sec:comp_model}
\begin{figure*}[thp]
\begin{center}
\includegraphics[width=0.98\textwidth]{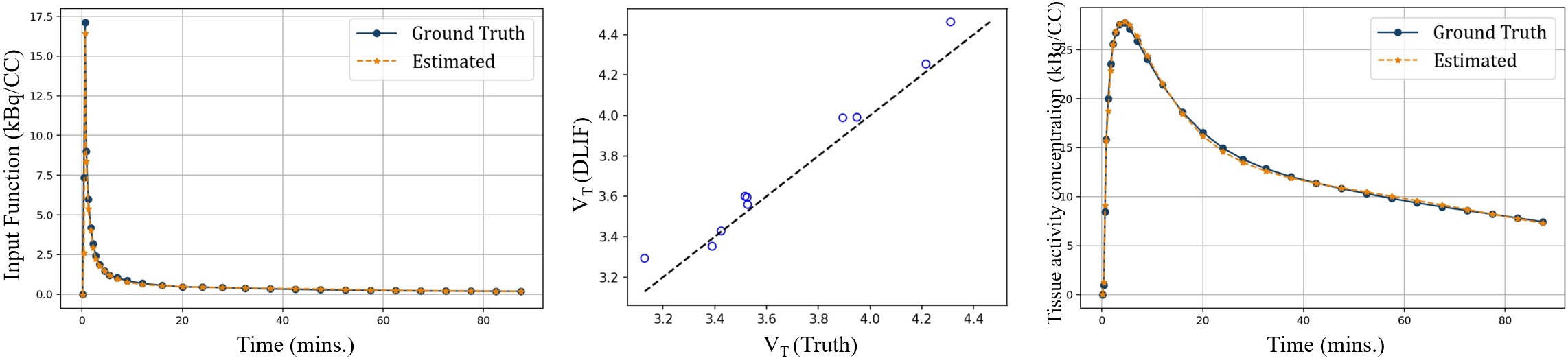}
\end{center}
   \caption{\textcolor{black}{Results of two-tissue compartment model analysis using the proposed DLIF. The left panel shows the estimated DLIF and the ground-truth AIF. The middle panel presents the correlation between total distribution volume ($V_T$) values derived using DLIF and the ground-truth AIF across different anatomical regions, where each circle in the plot represents a distinct region. The right panel illustrates the model fitting results for a representative region (frontal cortex) using both DLIF and ground-truth AIF.}}
\label{fig:two_fit}
\end{figure*}

\newpage
\section{List of Abbreviations}

\begin{table*}[tbh!]
\centering
\fontsize{10}{12}\selectfont
\caption{\textcolor{black}{List of abbreviations used throughout the paper. Acronyms are defined at first mention in the main text and summarized here for reference.}}
\label{tab:list_abbr}
\begin{tabular}{l l}
\Xhline{1pt}
\textbf{Abbreviation} & \textbf{Definition} \\
\Xhline{1pt}
\rowcolor{Gray}
AIF & Arterial input function (metabolite-corrected plasma input unless otherwise specified) \\
pAIF & Plasma arterial input function \\
 \rowcolor{Gray}
DLIF & Deep learning–derived input function \\
IDIF & Image-derived input function \\
 \rowcolor{Gray}
PBIF & Population-based input function \\
PET & Positron emission tomography \\
 \rowcolor{Gray}
CSF1R & Colony-stimulating factor 1 receptor \\
TSPO & Translocator protein, 18 kDa \\
\rowcolor{Gray}
$\text{[\textsuperscript{11}C]}$DPA-713 & TSPO-binding radiotracer \\
$\text{[\textsuperscript{11}C]}$CPPC & CSF1R-binding radiotracer \\
 \rowcolor{Gray}
HRRT & High Resolution Research Tomograph (Siemens ECAT HRRT scanner) \\
mCT & Siemens Biograph mCT PET/CT scanner \\
 \rowcolor{Gray}
ROI & Region of interest \\
PVE & Partial volume effect \\
 \rowcolor{Gray}
SUV & Standardized uptake value \\
TAC & Time–activity curve \\
 \rowcolor{Gray}
tTAC & Tissue time–activity curve \\
ODE & Ordinary differential equation \\
 \rowcolor{Gray}
DNN & Deep neural network \\
ConvNet & Convolutional neural network \\
 \rowcolor{Gray}
ViT & Vision Transformer \\
PINN & Physically informed neural network \\
 \rowcolor{Gray}
ICA & Independent component analysis \\
PCA & Principal component analysis \\
 \rowcolor{Gray}
NMF & Non-negative matrix factorization \\
EPICA & ICA-based IDIF estimation method~\citep{naganawa2005extraction, naganawa2005omission} \\
 \rowcolor{Gray}
EPISA & Extended projection or intersectional searching algorithm variant for 
IDIF~\citep{naganawa2008robust} \\
IoU & Intersection over Union (area overlap metric) \\
 \rowcolor{Gray}
PCC ($r$) & Pearson correlation coefficient \\
RMSE & Root mean squared error \\
 \rowcolor{Gray}
MAE & Mean absolute error \\
AUC & Area under the curve \\
 \rowcolor{Gray}
$V_T$ & Total distribution volume \\
HAB & High-affinity binder (TSPO genotype, C/C) \\
 \rowcolor{Gray}
MAB & Mixed-affinity binder (TSPO genotype, C/T) \\
LAB & Low-affinity binder (TSPO genotype, T/T) \\
\Xhline{1pt}
\end{tabular}
\end{table*}

\twocolumn
\bibliographystyle{model2-names.bst}
\biboptions{authoryear}
\bibliography{sample}

@article{naganawa2008robust,
  title={Robust estimation of the arterial input function for Logan plots using an intersectional searching algorithm and clustering in positron emission tomography for neuroreceptor imaging},
  author={Naganawa, Mika and Kimura, Yuichi and Yano, Junichi and Mishina, Masahiro and Yanagisawa, Masao and Ishii, Kenji and Oda, Keiichi and Ishiwata, Kiichi},
  journal={Neuroimage},
  volume={40},
  number={1},
  pages={26--34},
  year={2008},
  publisher={Elsevier}
}

@article{zanotti2011image,
  title={Image-derived input function for brain PET studies: many challenges and few opportunities},
  author={Zanotti-Fregonara, Paolo and Chen, Kewei and Liow, Jeih-San and Fujita, Masahiro and Innis, Robert B},
  journal={Journal of Cerebral Blood Flow \& Metabolism},
  volume={31},
  number={10},
  pages={1986--1998},
  year={2011},
  publisher={SAGE Publications Sage UK: London, England}
}

@article{naganawa2005omission,
  title={Omission of serial arterial blood sampling in neuroreceptor imaging with independent component analysis},
  author={Naganawa, Mika and Kimura, Yuichi and Nariai, Tadashi and Ishii, Kenji and Oda, Keiichi and Manabe, Yoshitsugu and Chihara, Kunihiro and Ishiwata, Kiichi},
  journal={Neuroimage},
  volume={26},
  number={3},
  pages={885--890},
  year={2005},
  publisher={Elsevier}
}

@article{naganawa2005extraction,
  title={Extraction of a plasma time-activity curve from dynamic brain PET images based on independent component analysis},
  author={Naganawa, Mika and Kimura, Yuichi and Ishii, Kenji and Oda, Keiichi and Ishiwata, Kiichi and Matani, Ayumu},
  journal={IEEE transactions on biomedical engineering},
  volume={52},
  number={2},
  pages={201--210},
  year={2005},
  publisher={IEEE}
}

@article{parker2006experimentally,
  title={Experimentally-derived functional form for a population-averaged high-temporal-resolution arterial input function for dynamic contrast-enhanced MRI},
  author={Parker, Geoff JM and Roberts, Caleb and Macdonald, Andrew and Buonaccorsi, Giovanni A and Cheung, Sue and Buckley, David L and Jackson, Alan and Watson, Yvonne and Davies, Karen and Jayson, Gordon C},
  journal={Magnetic Resonance in Medicine: An Official Journal of the International Society for Magnetic Resonance in Medicine},
  volume={56},
  number={5},
  pages={993--1000},
  year={2006},
  publisher={Wiley Online Library}
}

@article{chen1998noninvasive,
  title={Noninvasive quantification of the cerebral metabolic rate for glucose using positron emission tomography, 18F-fluoro-2-deoxyglucose, the Patlak method, and an image-derived input function},
  author={Chen, Kewei and Bandy, Daniel and Reiman, Eric and Huang, Sung-Cheng and Lawson, Michael and Feng, Dagan and Yun, Lang-sheng and Palant, Anita},
  journal={Journal of Cerebral Blood Flow \& Metabolism},
  volume={18},
  number={7},
  pages={716--723},
  year={1998},
  publisher={SAGE Publications Sage UK: London, England}
}

@article{su2005quantification,
  title={Quantification method in [18F] fluorodeoxyglucose brain positron emission tomography using independent component analysis},
  author={Su, Kuan-Hao and Wu, Liang-Chih and Liu, Ren-Shian and Wang, Shih-Jen and Chen, Jyh-Cheng},
  journal={Nuclear medicine communications},
  volume={26},
  number={11},
  pages={995--1004},
  year={2005},
  publisher={LWW}
}

@article{wahl1999regions,
  title={Regions of interest in the venous sinuses as input functions for quantitative PET},
  author={Wahl, Lindi M and Asselin, Marie-Claude and Nahmias, Claude},
  journal={Journal of Nuclear Medicine},
  volume={40},
  number={10},
  pages={1666--1675},
  year={1999},
  publisher={Soc Nuclear Med}
}

@article{litton1997input,
  title={Input function in PET brain studies using MR-defined arteries},
  author={Litton, Jan-Eric},
  journal={Journal of computer assisted tomography},
  volume={21},
  number={6},
  pages={907--909},
  year={1997},
  publisher={LWW}
}

@article{parker2005graph,
  title={Graph-based Mumford-Shah segmentation of dynamic PET with application to input function estimation},
  author={Parker, Brian J and Feng, Dagan},
  journal={IEEE Transactions on Nuclear Science},
  volume={52},
  number={1},
  pages={79--89},
  year={2005},
  publisher={IEEE}
}

@article{lee2012extraction,
  title={Extraction of an input function from dynamic micro-PET images using wavelet packet based sub-band decomposition independent component analysis},
  author={Lee, Jhih-Shian and Su, Kuan-Hao and Chang, Wen-Yuan and Chen, Jyh-Cheng},
  journal={NeuroImage},
  volume={63},
  number={3},
  pages={1273--1284},
  year={2012},
  publisher={Elsevier}
}

@article{mourik2008partial,
  title={Partial volume corrected image derived input functions for dynamic PET brain studies: methodology and validation for [11C] flumazenil},
  author={Mourik, Jurgen EM and Lubberink, Mark and Klumpers, Ursula MH and Comans, Emile F and Lammertsma, Adriaan A and Boellaard, Ronald},
  journal={Neuroimage},
  volume={39},
  number={3},
  pages={1041--1050},
  year={2008},
  publisher={Elsevier}
}

@article{liptrot2004cluster,
  title={Cluster analysis in kinetic modelling of the brain: a noninvasive alternative to arterial sampling},
  author={Liptrot, Matthew and Adams, Karen H and Martiny, Lars and Pinborg, Lars H and Lonsdale, Markus N and Olsen, Niels V and Holm, S{\o}ren and Svarer, Claus and Knudsen, Gitte M},
  journal={Neuroimage},
  volume={21},
  number={2},
  pages={483--493},
  year={2004},
  publisher={Elsevier}
}

@inproceedings{fung2009multimodal,
  title={A multimodal approach to image-derived input functions for brain PET},
  author={Fung, Edward K and Planeta-Wilson, Beata and Mulnix, Tim and Carson, Richard E},
  booktitle={2009 IEEE Nuclear Science Symposium Conference Record (NSS/MIC)},
  pages={2710--2714},
  year={2009},
  organization={IEEE}
}

@article{khalighi2018image,
  title={Image-derived input function estimation on a TOF-enabled PET/MR for cerebral blood flow mapping},
  author={Khalighi, Mohammad Mehdi and Deller, Timothy W and Fan, Audrey Peiwen and Gulaka, Praveen K and Shen, Bin and Singh, Prachi and Park, Jun-Hyung and Chin, Frederick T and Zaharchuk, Greg},
  journal={Journal of Cerebral Blood Flow \& Metabolism},
  volume={38},
  number={1},
  pages={126--135},
  year={2018},
  publisher={SAGE Publications Sage UK: London, England}
}

@article{su2013noninvasive,
  title={Noninvasive estimation of the arterial input function in positron emission tomography imaging of cerebral blood flow},
  author={Su, Yi and Arbelaez, Ana M and Benzinger, Tammie LS and Snyder, Abraham Z and Vlassenko, Andrei G and Mintun, Mark A and Raichle, Marcus E},
  journal={Journal of Cerebral Blood Flow \& Metabolism},
  volume={33},
  number={1},
  pages={115--121},
  year={2013},
  publisher={SAGE Publications Sage UK: London, England}
}

@article{fung2013cerebral,
  title={Cerebral blood flow with [15O] water PET studies using an image-derived input function and MR-defined carotid centerlines},
  author={Fung, Edward K and Carson, Richard E},
  journal={Physics in Medicine \& Biology},
  volume={58},
  number={6},
  pages={1903},
  year={2013},
  publisher={IOP Publishing}
}

@article{sari2017estimation,
  title={Estimation of an image derived input function with MR-defined carotid arteries in FDG-PET human studies using a novel partial volume correction method},
  author={Sari, Hasan and Erlandsson, Kjell and Law, Ian and Larsson, Henrik BW and Ourselin, Sebastien and Arridge, Simon and Atkinson, David and Hutton, Brian F},
  journal={Journal of Cerebral Blood Flow \& Metabolism},
  volume={37},
  number={4},
  pages={1398--1409},
  year={2017},
  publisher={SAGE Publications Sage UK: London, England}
}

@inproceedings{wang2006model,
  title={Model-based receptor quantization analysis for PET parametric imaging},
  author={Wang, Z Jane and Qiu, Peng and Liu, KJ Ray and Szabo, Zsolt},
  booktitle={2005 IEEE Engineering in Medicine and Biology 27th Annual Conference},
  pages={5908--5911},
  year={2006},
  organization={IEEE}
}

@article{logan1990graphical,
  title={Graphical analysis of reversible radioligand binding from time—activity measurements applied to [N-11C-methyl]-(-)-cocaine PET studies in human subjects},
  author={Logan, Jean and Fowler, Joanna S and Volkow, Nora D and Wolf, Alfred P and Dewey, Stephen L and Schlyer, David J and MacGregor, Robert R and Hitzemann, Robert and Bendriem, Bernard and Gatley, S John and others},
  journal={Journal of Cerebral Blood Flow \& Metabolism},
  volume={10},
  number={5},
  pages={740--747},
  year={1990},
  publisher={SAGE Publications Sage UK: London, England}
}

@article{b2006extraction,
  title={Extraction of time activity curves from positron emission tomography: K-means clustering or non-negative matrix factorization},
  author={Bödvarsson, Bjarni and Mørkebjerg, Martin and Hansen, Lars Kai and Knudsen, Gitte Moos and Svarer, Claus},
  journal={Neuroimage},
  volume={31},
  number={Suppl 1},
  pages={S154},
  year={2006},
  publisher={Elsevier Science}
}

@misc{wang2020direct,
  title={Direct estimation of input function based on fine-tuned deep learning method in dynamic pet imaging},
  author={Wang, Liangzhou and Ma, Tianyu and Yao, Shulin and Ye, Qing and Coughlin, Jennifer and Pomper, Martin and Du, Yong and Liu, Yaqiang},
  year={2020},
  publisher={Soc Nuclear Med}
}

@inproceedings{dosovitskiy2021an,
title={An Image is Worth 16x16 Words: Transformers for Image Recognition at Scale},
author={Alexey Dosovitskiy and Lucas Beyer and Alexander Kolesnikov and Dirk Weissenborn and Xiaohua Zhai and Thomas Unterthiner and Mostafa Dehghani and Matthias Minderer and Georg Heigold and Sylvain Gelly and Jakob Uszkoreit and Neil Houlsby},
booktitle={International Conference on Learning Representations},
year={2021}
}

@article{li2023transforming,
  title={Transforming medical imaging with Transformers? A comparative review of key properties, current progresses, and future perspectives},
  author={Li, Jun and Chen, Junyu and Tang, Yucheng and Wang, Ce and Landman, Bennett A and Zhou, S Kevin},
  journal={Medical image analysis},
  pages={102762},
  year={2023},
  publisher={Elsevier}
}

@article{vaswani2017attention,
  title={Attention is all you need},
  author={Vaswani, Ashish and Shazeer, Noam and Parmar, Niki and Uszkoreit, Jakob and Jones, Llion and Gomez, Aidan N and Kaiser, {\L}ukasz and Polosukhin, Illia},
  journal={Advances in neural information processing systems},
  volume={30},
  year={2017}
}

@inproceedings{liu2021swin,
  title={Swin transformer: Hierarchical vision transformer using shifted windows},
  author={Liu, Ze and Lin, Yutong and Cao, Yue and Hu, Han and Wei, Yixuan and Zhang, Zheng and Lin, Stephen and Guo, Baining},
  booktitle={Proceedings of the IEEE/CVF international conference on computer vision},
  pages={10012--10022},
  year={2021}
}

@article{han2022survey,
  title={A survey on vision transformer},
  author={Han, Kai and Wang, Yunhe and Chen, Hanting and Chen, Xinghao and Guo, Jianyuan and Liu, Zhenhua and Tang, Yehui and Xiao, An and Xu, Chunjing and Xu, Yixing and others},
  journal={IEEE transactions on pattern analysis and machine intelligence},
  volume={45},
  number={1},
  pages={87--110},
  year={2022},
  publisher={IEEE}
}

@article{endres2009initial,
  title={Initial evaluation of 11C-DPA-713, a novel TSPO PET ligand, in humans},
  author={Endres, Christopher J and Pomper, Martin G and James, Michelle and Uzuner, Ovsev and Hammoud, Dima A and Watkins, Crystal C and Reynolds, Aaron and Hilton, John and Dannals, Robert F and Kassiou, Michael},
  journal={Journal of Nuclear Medicine},
  volume={50},
  number={8},
  pages={1276--1282},
  year={2009},
  publisher={Soc Nuclear Med}
}

@article{coughlin2014regional,
  title={Regional brain distribution of translocator protein using [11 C] DPA-713 PET in individuals infected with HIV},
  author={Coughlin, Jennifer M and Wang, Yuchuan and Ma, Shuangchao and Yue, Chen and Kim, Pearl K and Adams, Ashley V and Roosa, Heidi V and Gage, Kenneth L and Stathis, Marigo and Rais, Rana and others},
  journal={Journal of neurovirology},
  volume={20},
  pages={219--232},
  year={2014},
  publisher={Springer}
}

@inproceedings{zhai2022scaling,
  title={Scaling vision transformers},
  author={Zhai, Xiaohua and Kolesnikov, Alexander and Houlsby, Neil and Beyer, Lucas},
  booktitle={Proceedings of the IEEE/CVF Conference on Computer Vision and Pattern Recognition},
  pages={12104--12113},
  year={2022}
}

@inproceedings{liu2022convnet,
  title={A convnet for the 2020s},
  author={Liu, Zhuang and Mao, Hanzi and Wu, Chao-Yuan and Feichtenhofer, Christoph and Darrell, Trevor and Xie, Saining},
  booktitle={Proceedings of the IEEE/CVF conference on computer vision and pattern recognition},
  pages={11976--11986},
  year={2022}
}

@inproceedings{ding2022scaling,
  title={Scaling up your kernels to 31x31: Revisiting large kernel design in cnns},
  author={Ding, Xiaohan and Zhang, Xiangyu and Han, Jungong and Ding, Guiguang},
  booktitle={Proceedings of the IEEE/CVF conference on computer vision and pattern recognition},
  pages={11963--11975},
  year={2022}
}

@inproceedings{liu2023more,
title={More ConvNets in the 2020s: Scaling up Kernels Beyond 51x51 using Sparsity},
author={Shiwei Liu and Tianlong Chen and Xiaohan Chen and Xuxi Chen and Qiao Xiao and Boqian Wu and Tommi K{\"a}rkk{\"a}inen and Mykola Pechenizkiy and Decebal Constantin Mocanu and Zhangyang Wang},
booktitle={The Eleventh International Conference on Learning Representations },
year={2023},
}

@inproceedings{he2016deep,
  title={Deep residual learning for image recognition},
  author={He, Kaiming and Zhang, Xiangyu and Ren, Shaoqing and Sun, Jian},
  booktitle={Proceedings of the IEEE conference on computer vision and pattern recognition},
  pages={770--778},
  year={2016}
}

@article{ba2016layer,
  title={Layer normalization},
  author={Ba, Jimmy Lei and Kiros, Jamie Ryan and Hinton, Geoffrey E},
  journal={arXiv preprint arXiv:1607.06450},
  year={2016}
}

@article{hoopes2022synthstrip,
  title={SynthStrip: Skull-stripping for any brain image},
  author={Hoopes, Andrew and Mora, Jocelyn S and Dalca, Adrian V and Fischl, Bruce and Hoffmann, Malte},
  journal={NeuroImage},
  volume={260},
  pages={119474},
  year={2022},
  publisher={Elsevier}
}

@article{mlynash2005automated,
  title={Automated method for generating the arterial input function on perfusion-weighted MR imaging: validation in patients with stroke},
  author={Mlynash, Michael and Eyngorn, Irina and Bammer, Roland and Moseley, Michael and Tong, David C},
  journal={American Journal of Neuroradiology},
  volume={26},
  number={6},
  pages={1479--1486},
  year={2005},
  publisher={Am Soc Neuroradiology}
}

@article{feng1993models,
  title={Models for computer simulation studies of input functions for tracer kinetic modeling with positron emission tomography},
  author={Feng, Dagan and Huang, Sung-Cheng and Wang, Xinmin},
  journal={International journal of bio-medical computing},
  volume={32},
  number={2},
  pages={95--110},
  year={1993},
  publisher={Elsevier}
}

@article{parsey2000validation,
  title={Validation and reproducibility of measurement of 5-HT1A receptor parameters with [carbonyl-11C] WAY-100635 in humans: comparison of arterial and reference tissue input functions},
  author={Parsey, Ramin V and Slifstein, Mark and Hwang, Dah-Ren and Abi-Dargham, Anissa and Simpson, Norman and Mawlawi, Osama and Guo, Ning-Ning and Van Heertum, Ronald and Mann, J John and Laruelle, Marc},
  journal={Journal of Cerebral Blood Flow \& Metabolism},
  volume={20},
  number={7},
  pages={1111--1133},
  year={2000},
  publisher={SAGE Publications Sage UK: London, England}
}

@inproceedings{
ferrante2022physically,
title={Physically Informed Neural Network for Non-Invasive Arterial Input Function Estimation In Dynamic {PET} Imaging},
author={Matteo Ferrante and Marianna Inglese and Ludovica Brusaferri and Alexander Whitehead and Marco Loggia and Nicola Toschi},
booktitle={Medical Imaging with Deep Learning},
year={2022}
}

@article{Chauveau_2008, year = {2008}, month = {oct}, pages = {2304--2319}, title = {Nuclear imaging of neuroinflammation: a comprehensive review of [11C]{PK}11195 challengers}, number = {12}, volume = {35}, journal = {European Journal of Nuclear Medicine and Molecular Imaging}, publisher = {Springer Science and Business Media {LLC}}, author = {Fabien Chauveau and Herv{\'{e}} Boutin and Nadja Van Camp and Fr{\'{e}}d{\'{e}}ric Doll{\'{e}} and Bertrand Tavitian}, }

@article{Venneti_2006, year = {2006}, month = {dec}, pages = {308--322}, title = {The peripheral benzodiazepine receptor (Translocator protein 18kDa) in microglia: From pathology to imaging}, number = {6}, volume = {80}, journal = {Progress in Neurobiology}, publisher = {Elsevier {BV}}, author = {Sriram Venneti and Brian J. Lopresti and Clayton A. Wiley}, }

@article{Tichauer_2015,
  title={Quantitative in vivo cell-surface receptor imaging in oncology: kinetic modeling and paired-agent principles from nuclear medicine and optical imaging},
  author={Tichauer, Kenneth M and Wang, Yu and Pogue, Brian W and Liu, Jonathan TC},
  journal={Physics in Medicine \& Biology},
  volume={60},
  number={14},
  pages={R239},
  year={2015},
  publisher={IOP Publishing}
}

@article{Muzi_2012, year = {2012}, month = {nov}, pages = {1203--1215}, title = {Quantitative assessment of dynamic {PET} imaging data in cancer imaging}, number = {9}, volume = {30}, journal = {Magnetic Resonance Imaging}, publisher = {Elsevier {BV}}, author = {Mark Muzi and Finbarr O{\textquotesingle}Sullivan and David A. Mankoff and Robert K. Doot and Larry A. Pierce and Brenda F. Kurland and Hannah M. Linden and Paul E. Kinahan}, }

@article{Wang_2016,
  title={Neuroimaging of translocator protein in patients with systemic lupus erythematosus: a pilot study using [11C] DPA-713 positron emission tomography},
  author={Wang, Yuchuan and Coughlin, Jennifer M and Ma, Shuangchao and Endres, Christopher J and Kassiou, Michael and Sawa, Akira and Dannals, Robert F and Petri, Michelle and Pomper, Martin G},
  journal={Lupus},
  volume={26},
  number={2},
  pages={170--178},
  year={2017},
  publisher={SAGE Publications Sage UK: London, England}
}

@article{Coughlin_2015, year = {2015}, month = {feb}, pages = {58--65}, title = {Neuroinflammation and brain atrophy in former {NFL} players: An in vivo multimodal imaging pilot study}, volume = {74}, journal = {Neurobiology of Disease}, publisher = {Elsevier {BV}}, author = {Jennifer M. Coughlin and Yuchuan Wang and Cynthia A. Munro and Shuangchao Ma and Chen Yue and Shaojie Chen and Raag Airan and Pearl K. Kim and Ashley V. Adams and Cinthya Garcia and Cecilia Higgs and Haris I. Sair and Akira Sawa and Gwenn Smith and Constantine G. Lyketsos and Brian Caffo and Michael Kassiou and Tomas R. Guilarte and Martin G. Pomper}, }

@article{Coughlin_2018, year = {2018}, month = {dec}, title = {Imaging glial activation in patients with post-treatment Lyme disease symptoms: a pilot study using [11C]{DPA}-713 {PET}}, number = {1}, volume = {15}, journal = {Journal of Neuroinflammation}, publisher = {Springer Science and Business Media {LLC}}, author = {Jennifer M. Coughlin and Ting Yang and Alison W. Rebman and Kathleen T. Bechtold and Yong Du and William B. Mathews and Wojciech G. Lesniak and Erica A. Mihm and Sarah M. Frey and Erica S. Marshall and Hailey B. Rosenthal and Tristan A. Reekie and Michael Kassiou and Robert F. Dannals and Mark J. Soloski and John N. Aucott and Martin G. Pomper}, }

@article{Takikawa_1993, year = {1993}, month = {jul}, pages = {131--136}, title = {Noninvasive quantitative fluorodeoxyglucose {PET} studies with an estimated input function derived from a population-based arterial blood curve.}, number = {1}, volume = {188}, journal = {Radiology}, publisher = {Radiological Society of North America ({RSNA})}, author = {S Takikawa and V Dhawan and P Spetsieris and W Robeson and T Chaly and R Dahl and D Margouleff and D Eidelberg}, }

@article{Zanotti_Fregonara_2012, year = {2012}, month = {nov}, pages = {1532--1541}, title = {Population-based input function and image-derived input function for [11C](R)-rolipram {PET} imaging: Methodology, validation and application to the study of major depressive disorder}, number = {3}, volume = {63}, journal = {{NeuroImage}}, publisher = {Elsevier {BV}}, author = {Paolo Zanotti-Fregonara and Christina S. Hines and Sami S. Zoghbi and Jeih-San Liow and Yi Zhang and Victor W. Pike and Wayne C. Drevets and Alan G. Mallinger and Carlos A. Zarate and Masahiro Fujita and Robert B. Innis}, }

@article{Boutin_2007, year = {2007}, month = {apr}, pages = {573--581}, title = {11C-{DPA}-713: A Novel Peripheral Benzodiazepine Receptor {PET} Ligand for In Vivo Imaging of Neuroinflammation}, number = {4}, volume = {48}, journal = {Journal of Nuclear Medicine}, publisher = {Society of Nuclear Medicine}, author = {H. Boutin and F. Chauveau and C. Thominiaux and M.-C. Gregoire and M. L. James and R. Trebossen and P. Hantraye and F. Dolle and B. Tavitian and M. Kassiou}, }

@article{Cui_2022,
year = {2022}, 
month = {aug},
pages = {102519}, 
title = {Unsupervised {PET} logan parametric image estimation using conditional deep image prior}, 
volume = {80}, 
journal = {Medical Image Analysis}, 
publisher = {Elsevier {BV}}, 
author = {Jianan Cui and Kuang Gong and Ning Guo and Kyungsang Kim and Huafeng Liu and Quanzheng Li}, }

@incollection{carson2005tracer,
  title={Tracer kinetic modeling in PET},
  author={Carson, Richard E},
  booktitle={Positron emission tomography: basic sciences},
  pages={127--159},
  year={2005},
  publisher={Springer}
}

@article{de2021aifnet,
  title={AIFNet: Automatic vascular function estimation for perfusion analysis using deep learning},
  author={de la Rosa, Ezequiel and Sima, Diana M and Menze, Bjoern and Kirschke, Jan S and Robben, David},
  journal={Medical Image Analysis},
  volume={74},
  pages={102211},
  year={2021},
  publisher={Elsevier}
}

@inproceedings{chen2023estimating,
  title={Estimating Arterial Input Function for Dynamic PET via Deep Regression},
  author={Chen, J and Jiang, Z and Coughlin, JM and Pomper, MG and Du, Y},
  booktitle={2023 IEEE Nuclear Science Symposium, Medical Imaging Conference and International Symposium on Room-Temperature Semiconductor Detectors (NSS MIC RTSD)},
  pages={1--1},
  year={2023},
  organization={IEEE}
}

@article{paszke2019pytorch,
  title={Pytorch: An imperative style, high-performance deep learning library},
  author={Paszke, Adam and Gross, Sam and Massa, Francisco and Lerer, Adam and Bradbury, James and Chanan, Gregory and Killeen, Trevor and Lin, Zeming and Gimelshein, Natalia and Antiga, Luca and others},
  journal={Advances in neural information processing systems},
  volume={32},
  year={2019}
}

@article{kingma2014adam,
  title={Adam: A method for stochastic optimization},
  author={Kingma, Diederik P and Ba, Jimmy},
  journal={arXiv preprint arXiv:1412.6980},
  year={2014}
}

@article{Owen201118kDa,
	journal = {Journal of Cerebral Blood Flow and Metabolism},

	title = {An 18-kDa Translocator Protein (TSPO) Polymorphism Explains Differences in Binding Affinity of the PET Radioligand PBR28},
	volume = {32},
	author = {Owen, David R and Yeo, Astrid J and Gunn, Roger N and Song, Kijoung and Wadsworth, Graham and Lewis, Andrew and Rhodes, Chris and Pulford, David J and Bennacef, Idriss and Parker, Christine A and StJean, Pamela L and Cardon, Lon R and Mooser, Vincent E and Matthews, Paul M and Rabiner, Eugenii A and Rubio, Justin P},
	pages = {1--5},
	year = {2011},
	month = {10},
	day = {19},
}

@article{Milenkovic2018Effects,
	journal = {PLOS ONE},
	title = {Effects of genetic variants in the TSPO gene on protein structure and stability},
	volume = {13},
	author = {Milenkovic, Vladimir M. and Bader, Stefanie and Sudria-Lopez, Daniel and Siebert, Ramona and Brandl, Caroline and Nothdurfter, Caroline and Weber, Bernhard H. F. and Rupprecht, Rainer and Wetzel, Christian H.},
	pages = {e0195627},
	year = {2018},
	month = {4},
	day = {11},
}

@article{dimitrakopoulou2021kinetic,
  title={Kinetic modeling and parametric imaging with dynamic PET for oncological applications: general considerations, current clinical applications, and future perspectives},
  author={Dimitrakopoulou-Strauss, Antonia and Pan, Leyun and Sachpekidis, Christos},
  journal={European journal of nuclear medicine and molecular imaging},
  volume={48},
  pages={21--39},
  year={2021},
  publisher={Springer}
}

@article{rahmim2019dynamic,
  title={Dynamic whole-body PET imaging: principles, potentials and applications},
  author={Rahmim, Arman and Lodge, Martin A and Karakatsanis, Nicolas A and Panin, Vladimir Y and Zhou, Yun and McMillan, Alan and Cho, Steve and Zaidi, Habib and Casey, Michael E and Wahl, Richard L},
  journal={European journal of nuclear medicine and molecular imaging},
  volume={46},
  pages={501--518},
  year={2019},
  publisher={Springer}
}

@article{gunn2001positron,
  title={Positron emission tomography compartmental models},
  author={Gunn, Roger N and Gunn, Steve R and Cunningham, Vincent J},
  journal={Journal of Cerebral Blood Flow \& Metabolism},
  volume={21},
  number={6},
  pages={635--652},
  year={2001},
  publisher={SAGE Publications Sage UK: London, England}
}

@article{watabe2006pet,
  title={PET kinetic analysis—compartmental model},
  author={Watabe, Hiroshi and Ikoma, Yoko and Kimura, Yuichi and Naganawa, Mika and Shidahara, Miho},
  journal={Annals of nuclear medicine},
  volume={20},
  pages={583--588},
  year={2006},
  publisher={Springer}
}

@article{coughlin2018distribution,
  title={The distribution of the alpha7 nicotinic acetylcholine receptor in healthy aging: an in vivo positron emission tomography study with [18F] ASEM},
  author={Coughlin, Jennifer M and Du, Yong and Rosenthal, Hailey B and Slania, Stephanie and Koo, Soo Min and Park, Andrew and Solomon, Ghedem and Vranesic, Melin and Antonsdottir, Inga and Speck, Caroline L and others},
  journal={Neuroimage},
  volume={165},
  pages={118--124},
  year={2018},
  publisher={Elsevier}
}

@article{kang2018noninvasive,
  title={Noninvasive PK11195-PET image analysis techniques can detect abnormal cerebral microglial activation in Parkinson's disease},
  author={Kang, Yeona and Mozley, P David and Verma, Ajay and Schlyer, David and Henchcliffe, Claire and Gauthier, Susan A and Chiao, Ping C and He, Bin and Nikolopoulou, Anastasia and Logan, Jean and others},
  journal={Journal of Neuroimaging},
  volume={28},
  number={5},
  pages={496--505},
  year={2018},
  publisher={Wiley Online Library}
}

@article{rubin2018microglial,
  title={Microglial activation is inversely associated with cognition in individuals living with HIV on effective antiretroviral therapy},
  author={Rubin, Leah H and Sacktor, Ned and Creighton, Jason and Du, Yong and Endres, Christopher J and Pomper, Martin G and Coughlin, Jennifer M},
  journal={Aids},
  volume={32},
  number={12},
  pages={1661--1667},
  year={2018},
  publisher={LWW}
}

@article{rubin2023imaging,
  title={Imaging brain injury in former National Football League players},
  author={Rubin, Leah H and Du, Yong and Sweeney, Shannon Eileen and O’Toole, Riley and Thomas, Cykyra L and Zandi, Adeline G and Shinehouse, Laura K and Brosnan, Mary Katherine and Nam, Hwanhee and Burke, Michael E and others},
  journal={JAMA Network Open},
  volume={6},
  number={10},
  pages={e2340580--e2340580},
  year={2023},
  publisher={American Medical Association}
}

@article{rubin2022imaging,
  title={Imaging the translocator protein 18 kDa within cognitive control and declarative memory circuits in virally-suppressed people with HIV},
  author={Rubin, Leah H and Maki, Pauline M and Du, Yong and Sweeney, Shannon Eileen and O’toole, Riley and Nam, Hwanhee and Lee, Hannah and Soule, Ana R and Rowe, Steven P and Lesniak, Wojciech G and others},
  journal={AIDS},
  pages={10--1097},
  year={2022},
  publisher={LWW}
}

@article{ferrante2024physically,
  title={Physically informed deep neural networks for metabolite-corrected plasma input function estimation in dynamic PET imaging},
  author={Ferrante, Matteo and Inglese, Marianna and Brusaferri, Ludovica and Whitehead, Alexander C and Maccioni, Lucia and Turkheimer, Federico E and Nettis, Maria A and Mondelli, Valeria and Howes, Oliver and Loggia, Marco L and others},
  journal={Computer methods and programs in biomedicine},
  volume={256},
  pages={108375},
  year={2024},
  publisher={Elsevier}
}

@article{coughlin2022first,
  title={First-in-human use of 11C-CPPC with positron emission tomography for imaging the macrophage colony-stimulating factor 1 receptor},
  author={Coughlin, Jennifer M and Du, Yong and Lesniak, Wojciech G and Harrington, Courtney K and Brosnan, Mary Katherine and O’Toole, Riley and Zandi, Adeline and Sweeney, Shannon Eileen and Abdallah, Rehab and Wu, Yunkou and others},
  journal={EJNMMI research},
  volume={12},
  number={1},
  pages={64},
  year={2022},
  publisher={Springer}
}

@article{horti2019pet,
  title={PET imaging of microglia by targeting macrophage colony-stimulating factor 1 receptor (CSF1R)},
  author={Horti, Andrew G and Naik, Ravi and Foss, Catherine A and Minn, IL and Misheneva, Varia and Du, Yong and Wang, Yuchuan and Mathews, William B and Wu, Yunkou and Hall, Andrew and others},
  journal={Proceedings of the National Academy of Sciences},
  volume={116},
  number={5},
  pages={1686--1691},
  year={2019},
  publisher={National Academy of Sciences}
}

@article{mills2025exploring,
  title={Exploring [11 C] CPPC as a CSF1R-targeted PET imaging marker for early Parkinson’s disease severity},
  author={Mills, Kelly A and Du, Yong and Coughlin, Jennifer M and Foss, Catherine A and Horti, Andrew G and Jenkins, Katelyn R and Skorobogatova, Yana and Spiro, Ergi and Motley, Chelsie S and Dannals, Robert F and others},
  journal={The Journal of Clinical Investigation},
  year={2025},
  publisher={American Society for Clinical Investigation}
}

@article{kuttner2024deep,
  title={Deep-learning-derived input function in dynamic [18F] FDG PET imaging of mice},
  author={Kuttner, Samuel and Luppino, Luigi T and Convert, Laurence and Sarrhini, Otman and Lecomte, Roger and Kampffmeyer, Michael C and Sundset, Rune and Jenssen, Robert},
  journal={Frontiers in Nuclear Medicine},
  volume={4},
  pages={1372379},
  year={2024},
  publisher={Frontiers Media SA}
}

\end{document}